\documentclass[aps,pre,preprint,floatfix]{revtex4}
\usepackage[utf8]{inputenc}
\usepackage{amsmath}
\usepackage{graphicx}
\usepackage{amssymb}
\usepackage{epsfig}
\usepackage{natbib}

\newcommand{\BE}{\begin{equation}}
\newcommand{\EE}{\end{equation}}

\usepackage{graphicx}% Include figure files
\usepackage{dcolumn}% Align table columns on decimal point
\usepackage{bm}% bold math
\def\bq{\begin{equation}}
\def\eq{\end{equation}}

\begin{document}

%\preprint{APS/123-QED}

%%%%%%%%%%%%%%%%%%%%%%%%%%%%%%%%%%%%%%%%
\title{Front propagation in reaction-diffusion systems \\ with anomalous diffusion}
%%%%%%%%%%%%%%%%%%%%%%%%%%%%%%%%%%%%%%%%

%\author{dCN}
%\author{Diego \& Luis}
\author{D. del-Castillo-Negrete}
\email{delcastillod@ornl.gov}
\affiliation{Oak Ridge National Laboratory \\ Oak Ridge TN, 37831-8071}
\date{\today}% It is always \today, today,

\begin{abstract}
A numerical study of the role of anomalous diffusion in front propagation in reaction-diffusion systems is presented. Three models of anomalous diffusion are considered: fractional diffusion, tempered fractional diffusion, and a model that combines  fractional diffusion and regular diffusion. The reaction kinetics corresponds to a Fisher-Kolmogorov nonlinearity.
The numerical method is based on a finite-difference operator splitting algorithm with an explicit Euler step for the time advance of the reaction kinetics, and a Crank-Nicholson semi-implicit time step for the transport operator. The anomalous diffusion operators are discretized using an upwind,  flux-conserving,  Grunwald-Letnikov finite-difference scheme applied to the regularized fractional derivatives.
With fractional diffusion of order $\alpha$, fronts exhibit exponential acceleration, $a_L(t) \sim e^{\gamma t/\alpha}$, and develop algebraic decaying tails, $\phi \sim 1/x^{\alpha}$.  In the case of tempered fractional diffusion, this phenomenology prevails in the intermediate asymptotic regime
 $\left(\chi t \right)^{1/\alpha} \ll x \ll 1/\lambda$, where $1/\lambda$ is the scale of the tempering. Outside this regime, i.e. for $x > 1/\lambda$, the tail exhibits the tempered decay 
$\phi \sim e^{-\lambda x}/x^{\alpha+1}$,  and the front velocity approaches the terminal speed $v_*= \left(\gamma-\lambda^\alpha \chi\right)/ \lambda$.
Of particular interest is the study of the interplay of regular and fractional diffusion.
It is shown that the main role of regular diffusion is to delay the onset of front acceleration.  In particular, the crossover time, $t_c$, to transition to the accelerated fractional regime exhibits a logarithmic scaling of the form $t_c \sim \log \left(\chi_d/\chi_f\right)$ where $\chi_d$ and $\chi_f$ are the regular and fractional diffusivities.
\end{abstract}

%\pacs{52.65.-y, 52.25.Dg, 52.25.Fi, 02.70.Ns, 02.70.Uu}

% 52.65.-y 	Plasma simulation
% 52.25.Dg 	Plasma kinetic equations
% 52.25.Fi 	Transport properties 
%02.70.Ns 	Molecular dynamics and particle methods
% 02.70.Uu 	Applications of Monte Carlo methods 

\maketitle

%%%%%%%%%%%%%%%%%%%%%%%%
\section{Introduction}
%%%%%%%%%%%%%%%%%%%%%%%%

Front propagation is a problem of interest to many areas including biology, plasma physics, fluid dynamics, chemistry, materials, and engineering.  In the widely studied  case of reaction diffusion systems, front propagation arises from the coupling of transport with the reaction kinetics.  Among the simplest systems of this type is the 
Fisher-Kolmogorov model that describes the dynamics of a 
scalar field, $\phi$, in a one-dimensional domain,
\bq
\label{fk}
\partial_t \phi = \chi_d \partial_x^2 \phi + \gamma \phi \left ( 1 - 
\phi \right) \, ,
\eq
where $\chi_d$ denotes the diffusivity, and $\gamma$ is a constant \cite{rd_biology}. 
In this case, the competition of diffusion and the nonlinear reaction, leads to the 
propagation of fronts in which the stable, $\phi=1$, state advances 
through the destabilization of the $\phi=0$ unstable state.  

An implicit assumption in the majority of reaction-diffusion models  is that transport satisfies the Fourier-Fick's 
prescription according to which the flux, $q$, of the transported field, $\phi$, is proportional to the gradient, i.e.,
\bq
\label{ficks}
q = - \chi_d \partial_x \phi \, .
\eq
Substituting Eq.~(\ref{ficks}) in  the conservation law, $\partial_t \phi = -\partial_x q$, gives the well-known diffusive transport model, $\partial_t \phi =\chi \partial^2_x \phi$. 
From the statistical mechanics point of view, the Fourier-Fick's 
prescription is closely related to the assumption that the underlying 
``microscopic'' transport process is driven by a Markovian, Gaussian process.  
Although this assumption has proved to be valuable, it fails to apply in the case of anomalous transport. 

Anomalous transport naturally arises in systems in which the underlying statistics is non-Gaussian. A paradigmatic example 
is the problem of transport in fluids in the presence of coherent structures, e.g., vortices and zonal flows. In this case, 
experimental \cite{solomon} and theoretical studies \cite{del_castillo_1998} have shown that the trapping effect of vortices and the 
long displacements induced by zonal flows leads to non-Gaussian Lagrangian statistics and non-diffusive transport.
A similar phenomenology has also been observed in studies of transport in plasma turbulence \cite{del_castillo_2005}. 
From a statistical mechanics perspective, transport in these systems can be studied in the framework of the continuous time random walk model that generalizes the standard Brownian random walk by incorporating general waiting time distributions (to describe non-Markovian, ``memory" effects) and  L\'evy flights (to describe anomalously large jumps) \cite{metzler_2000}. 
At the macroscopic level, it turns out that in this type of systems, Eq.~(\ref{ficks}) is not satisfied because 
the flux at a given point in space depends on the gradient of the concentration throughout the entire domain. That is, the flux-gradient relation exhibits a nonlocal dependence of the form 
\bq
\label{non_local_q}
q = - \chi \, \partial_x \int {\cal K}(x-x') \phi(x') \, dx' \, ,
\eq
where the kernel, ${\cal K}(x-x')$, determines the degree of non-locality.
For an application of this type of non-local models to non-diffusive transport experiments in plasma physics see Ref.~\cite{dcn_nf_2008}. A general, tutorial-type  discussion on these  models can be found in Refs.~\cite{iter,book_2011}. 

The goal of the present paper is to address the problem of non-local transport in the context of front propagation.  
Our study is based on different generalizations of the Fisher-Kolmogorov model in Eq.~(\ref{fk}) in which the  standard diffusion operator, $\partial^2_x \phi$,  is substituted by non-local operators of the form in 
Eq.~(\ref{non_local_q}) describing different aspect of anomalous transport.  Among the most remarkable features of front propagation in the presence of non-local  transport is front acceleration and the development of algebraic decaying tails. These properties were originally discussed in Ref.~\cite{del_castillo_2003} in the context of the non-local, fractional Fisher-Kolmogorov model, and in the context of a master equation probabilistic model in Ref.~\cite{vulpiani}. 
The role of tempered L\'evy processes in front propagation was discussed in 
Ref.~\cite{del_castillo_2009}, where it was shown that the truncation of the L\'evy flights due to tempering, leads to a transient front acceleration after which the front asymptotically reaches a terminal speed. 
Other works on anomalous transport in front propagation include studies on: 
bistable reaction processes and anomalous diffusion caused by L\'evy flights \cite{zanette_1997}; 
analytic solutions of fractional reaction-diffusion equations
\cite{saxena_sols_2006}; reaction-diffusion systems with bistable 
reaction terms and directional anomalous diffusion \cite{barrio_2006}; 
construction of 
reaction-sub-diffusion equations \cite{sokolov_fronts_2006};  
fractional reproduction-dispersal equations and heavy tail dispersal kernels 
\cite{Baeumer_etal_2007}; role of fluctuations in reaction-super-diffusion dynamics
\cite{brockman};
non-Markovian random walks and sub-diffusion in reaction-diffusion systems \cite{fedotov_2010};
exact super-diffusive front propagation solutions with piecewise linear reaction kinetics functions
\cite{volpert_etal_2010}; and front dynamics in two-species competition models driven by L\'evy flights \cite{hanert_2012}, among others.  
The recent work in Ref.~\cite{cabre_2013} discusses a mathematically rigorous justification of 
some of the results on nonlocal front propagation originally presented in Refs.~\cite{del_castillo_2003,vulpiani}.

The organization of the rest of the paper is as follows. 
In the next section we present a brief overview of the different models of anomalous transport of interest in the present paper. In Sec.~II we study front propagation in the presence of the anomalous transport  operators discussed in Sec.~I. A summary of the results and the conclusions are presented in Sec.~III. 

%%%%%%%%%%%%%%%%%%%%%%%%
\section{Anomalous diffusion}
%%%%%%%%%%%%%%%%%%%%%%%%

Our strategy to study the role of anomalous transport in front propagation is based on modifications of the Fisher-Kolmogorov equation in which the standard second derivative diffusion operator is changed to nonlocal operators capable of modeling different aspects of anomalous transport. 
In this section, following a brief description of the main properties of the diffusion operator, we present a brief review of the basic properties of the nonlocal operators of interest. Further details can be found in 
Refs.~\cite{iter,book_2011} and references therein. 

%%%%%%%%%%%%%%%%%%%%%
\subsection{Standard diffusion}
%%%%%%%%%%%%%%%%%%%%%

The starting point to model transport of a single scalar field, $\phi(x,t)$, in a one-dimensional domain, is the conservation law,
%\bq
%\frac{\partial \phi}{\partial t} = -\frac{\partial q}{\partial x} + S \, ,
%\eq
\bq
\label{conserv}
\partial_t \phi= - \partial_x q + S \, ,
\eq
where $q(x,t)$ is the flux, and $S(x,t)$ is the source. Within the standard diffusion paradigm, the flux is modeled using the Fourier-Fick's prescription,
\bq
\label{fick}
q = - \chi_d \partial_x \phi + V_d \phi \, ,
\eq
where $\chi_d$ is the diffusivity, and $V_d$ the drift velocity. 
Substituting Eq.~(\ref{fick})  into Eq.~(\ref{conserv}) leads to the advection-diffusion equation,
\bq
\label{adv_diff}
%{\cal D}\left[\phi \right]=\chi_d \partial_x^2 \phi \, ,
\partial_t \phi +  \partial_x \left( V_d \phi \right) =\partial_x \left( \chi_d  \partial_x \phi \right) +S \, .
\eq
For $S=0$, $V_d=0$, and $\chi_d$ constant, Eq.~(\ref{adv_diff})
reduces to the  diffusion equation, 
\bq
\label{diff}
%{\cal D}\left[\phi \right]=\chi_d \partial_x^2 \phi \, ,
\partial_t \phi=\chi_d \partial_x^2 \phi \,  .
\eq
As it is well-known, the solution of the initial value problem of Eq.~(\ref{diff}) with  $\phi_0(x)=\phi(x,t=0)$ and prescribed boundary conditions, is given by
\bq
\label{ivp}
\phi(x,t)= \int \phi_0(x') G(x-x',t) dx'  \, ,
\eq
where $G$ is the Green's function for the corresponding boundary conditions. 
In an unbounded domain, $(-\infty, \infty)$,  $G$ is given by
%the Gaussian distribution 
%$$
%G(x,t)=\frac{1}{\sqrt{4 \pi \chi_d t }} \exp\left[ -\frac{x^2}{4 \chi_d t}\right] \, . 
%$$
%This solution can be obtained by looking for self-similar solutions of the form
\bq
\label{similarity}
G(x,t)=\left( \chi t \right)^{-\gamma/2} L(\eta) \, , \qquad 
\eta = \frac{x}{\left( \chi t \right)^{\gamma/2}}\, ,
\eq
where the scaling function, $L$, and the scaling exponent, $\gamma$, are 
\bq
L(\eta)=\frac{1}{2 \sqrt{\pi}} e^{-\eta^2/4}\, , \qquad \gamma=1 \, .
\eq
From Eq.~(\ref{similarity}) it follows that the time evolution of the moments of order $q$ scales as,
\bq
\label{moments}
%\langle \left[ \delta x -\langle x \rangle \right] ^n \rangle \sim t^{n \gamma}  
\langle  x^q \rangle \sim t^{q \gamma/2}  
\eq
where $\langle g \rangle = \int g(x) G(x,t)\, dx$. In particular, the second moment, $q=2$,  grows linearly with time. 

%%%%%%%%%%%%%%%%%%%%%
\subsection{Fractional diffusion}
%%%%%%%%%%%%%%%%%%%%%

In the fractional diffusion model, 
\bq
\label{frac}
\partial_t \phi =\chi_f \left[ l \,
_{a}D_x^\alpha +  r \, _{x}D_{b}^\alpha \right] \, \phi \, ,\eq
where $x \in (a,b)$, and $1<\alpha <2$.
The operators on the right hand side of Eq.(\ref{frac}) are the left and right 
Riemann-Liouville fractional derivatives
\cite{podlu_1999,samko_1993},
\bq
\label{left}
_{a}D_x^\alpha \phi= \frac{1}{\Gamma(m-\alpha)}\,
\frac{\partial^m}{\partial x^m}\,
\int_a^x\, \frac{\phi}{(x-y)^{\alpha+1-m}}\, dy \, ,
\eq
\bq
\label{right}
_{x}D_b^\alpha \phi = \frac{(-1)^m}{\Gamma(m-\alpha)}\,
\frac{\partial^m}{\partial x^m}\,
\int_x^b\, \frac{\phi}{(y-x)^{\alpha+1-m}}\, dy \, ,
\eq
with $m-1 \leq \alpha < m$, and 
the weighting factors $l$ and $r$ are defined as
\begin{equation}
\label{lr}
l=-\frac{(1-\theta)}{2 \cos(\alpha \pi/2)}\, , \qquad
r=-\frac{(1+\theta)}{2 \cos(\alpha \pi/2)} \, ,
\end{equation}
with $-1\leq \theta \leq1$.
Equation~(\ref{frac}) can be written in the flux-conserving form in
Eq.~(\ref{conserv}) with $q=q_l + q_r$ where
\bq
\label{l_flux}
q_l=- l \chi _l \,  _aD_x^{\alpha-1}\,  \phi\, ,
\qquad
%\label{r_flux}
q_r= r \chi _r \,  _xD_b^{\alpha-1}\, \phi \, .
%\qquad q_g = - \chi_d \, \partial_x P \, .
\eq
Using 
\bq
{\cal F} \left[ _{-\infty}D_x^{\alpha} \phi \right] =\left(-i k \right)^\alpha \hat{\phi} \, , \qquad
{\cal F} \left[ _{x}D_\infty^{\alpha} \phi \right] =\left(i k \right)^\alpha \hat{\phi} \, ,
\eq
where ${\cal F}[\phi]=\hat \phi$ denotes the Fourier transform, it follows that in an unbounded domain the solution of the initial value problem of Eq.~(\ref{frac}), is Eq.~(\ref{ivp}) with the Green function given in Eq.~(\ref{similarity}) with 
%\bq
%\label{green_fcn}
%G(x,t)=t^{-1/\alpha} \, L (\eta )\, , 
%\eq
\bq
\label{k_fcn}
L(\eta)= \frac{1}{2 \pi} \int_{-\infty}^{\infty} e^{-i \eta k+ \Lambda (k)}\,
d k\, , \qquad \gamma=2/\alpha  \, .
\eq
The scaling function, $L$, is the $\alpha$-stable L\'evy distribution,
with characteristic exponent
\bq
\label{psi_kk}
\Lambda =\chi_f \left[ l (-ik)^\alpha + r(ik)^\alpha \right]\, . % -\chi_d k^2 \, ,
\eq
There are two key properties of the Green's function of the fractional diffusion equation. One is the algebraic decay of the tails, 
\bq
\label{x_alpha_scaling}
G (x,t)\sim x^{-(1+\alpha)} \, , \qquad x\gg \left( \chi_f
t \right)^{1/\alpha}
\, ,
\eq
which follows from the asymptotic expansion of $L(\eta)$ in Eq.~(\ref{k_fcn}) for $\eta = x/\left( \chi_f
t \right)^{1/\alpha}\gg 1$. 
The other key property is the anomalous scaling of the moments, which, according to
Eq.~(\ref{moments}),  exhibit super-diffusive growth for $\alpha <2$. Note however that, in this case, only the moments with $q <2$ exist.  

%%%%%%%%%%%%%%%%%%%%%
\subsection{Tempered fractional diffusion}
%%%%%%%%%%%%%%%%%%%%%

Although the existence of large displacements has been documented in experimental and theoretical studies of super-diffusive transport, in applications it is clear that  displacements cannot be arbitrarily large. That is, strictly speaking, the concept of L\'evy flights (i.e., random displacements with infinite second moments)  is a mathematical idealization. It is thus of significant interest to construct models of anomalous transport that incorporate large events while keeping the moments finite. 
From the statistical mechanics perspective, the key issue is to temper the L\'evy flights by introducing a truncation length scale. 
%Truncated L\'evy distributions were originally introduced in Ref.~\cite{mantegna,kopone} as a natural prescription to guarantee the finiteness of the second moments. 
This problem was addressed in Ref.~\cite{cartea_del_castillo_2007} where a 
new model of anomalous transport  was constructed as the macroscopic limit of a continuous time random walk model with jumps following an exponentially tempered L\'evy process. 
The model, which describes  the interplay of long jumps and truncation in the intermediate asymptotic regime, has the form 
\begin{equation}
\label{trunc}
\partial_t \phi = -V \partial_x \phi +
\chi_t {\cal D}_x^{\alpha,\lambda} \phi -\mu \phi   \,  , 
\end{equation}
where the $\lambda$-truncated fractional derivative operator of order $\alpha$,
${\cal D}_x^{\alpha,\lambda}$,
is defined as~\cite{cartea_del_castillo_2007}
\begin{equation}
\label{temp_1}
{\cal D}_x^{\alpha,\lambda} = l e^{-\lambda x}\,
_{-\infty}D_x^\alpha\, e^{\lambda x}\, +r e^{\lambda
x}\, _{x}D_{\infty}^\alpha\, e^{-\lambda x}  \, .
\end{equation}
The effective advection velocity is  $V=V_d$ for $0<\alpha<1$, and $V=V_d-v$ for  $1<\alpha<2$ where 
\bq
\label{temp_2} 
v=  \frac{\chi_t \alpha \theta \lambda^{\alpha-1}}{\left|
\cos\left( \alpha \pi /2\right)\right|} \, , \qquad  \mu = -\frac{\chi_t \lambda^\alpha}{\cos\left(
\alpha \pi /2\right)}\, ,
\end{equation} 
and $V_d$ denotes a constant background drift velocity which is independent of the specific diffusive transport considered. 
The term $v$ results from the biasing of the tempered L\'evy process and it vanishes in the symmetric, $\theta=0$, case.

Using
\begin{equation}
\label{eq_26_1}
{\cal F}\left [
  e^{-\lambda x}\, _{-\infty}D_x^\alpha\, e^{ \lambda x}\, \phi
\right]= \left(\lambda - ik \right)^\alpha \, \hat{\phi}\, , \qquad
{\cal F}\left [
  e^{\lambda x}\, _{x}D_\infty^\alpha\, e^{ -\lambda x}\, \phi
\right]= \left(\lambda + ik \right)^\alpha \, \hat{\phi}\,
\end{equation}
it  follows that the Green's function of Eq.~(\ref{trunc}) with $1<\alpha < 2$ is 
\begin{equation}
\label{eq_17_2}
G(x,t)=\frac{1}{2\pi} \int_{-\infty}^\infty e^{-ik x+t \Lambda} dk \, ,
\end{equation}
where the characteristic exponent, $\Lambda$, is
%\bq
%\Lambda=\frac{-\chi_t}{2 \cos(\alpha\pi/2)}  \left[
%(1+\theta)(\lambda +ik )^{\alpha }+
%(1-\theta)(\lambda -ik )^{\alpha }-
%2 \lambda^{\alpha }- 2 i k\alpha \theta \lambda ^{\alpha -1}  \right]\, .
%\eq
\bq
\label{ce}
\Lambda=\frac{-\chi_t}{2 \cos(\alpha\pi/2)}  \left[
(1+\theta)(\lambda +ik )^{\alpha }+
(1-\theta)(\lambda -ik )^{\alpha }  \right] - i k v - \mu
 \, .
\eq
As mentioned before, the parameter, $-1 < \theta < 1$,  determines the asymmetry of the stochastic process. 

The interpretation of the parameter $\mu$ in Eqs.~(\ref{trunc}) and  (\ref{ce}) deserves some discussion. Eventhough the term 
``$-\mu \phi$"  looks like a ``damping", its role is exactly the opposite. In particular, this term (which follows from the derivation of 
 Eq.~(\ref{trunc}) from the continuous time random walk model \cite{cartea_del_castillo_2007}) is critical to guarantee the conservation of $\phi$, that is, to guarantee that $d M /dt =0$, where $M=\int_{-\infty}^\infty \phi dx$. Imposing this conservation law is fundamental because, in an unbounded domain, regular and anomalous transport processes must conserve the total ``mass" (or whatever physical property the integral of $\phi$ represents). To see this, note that according to Eq.~(\ref{ivp}), $M(t)=\hat{G}(k=0,t) M(0)$, where we have applied the convolution theorem and used  $M=\hat{\phi}(k=0,t)$. From Eq.~(\ref{eq_17_2}) it follows that
 $\hat{G}(k=0,t)=e^{t \Lambda(0)}$, and thus $M$ will be conserved if and only if $\Lambda(0)=0$, which according to  Eq.~(\ref{ce}), 
is guaranteed by the ``$-\mu$" term. 

One of the key properties of the tempered fractional diffusion Eq.~(\ref{trunc})  is 
the finiteness of all the moments of the Green's function. In particular, in the case $V=0$,
the second moment exhibit the diffusive scaling
\begin{equation}
\label{eq_33}
\left \langle  x^2
\right \rangle \sim 
\frac{\chi_t \alpha |\alpha-1| }{2|\cos(\alpha \pi /2)|
\lambda^{2-\alpha}} \, t \, ,
\end{equation} 
for large $t$. 
The truncation has also nontrivial effects on the  tails of the Green's function which 
exhibit the exponentially tempered decay,
\bq
\label{temp_decay}
G(x,t_0)  \sim \frac{ e^{-\lambda x}}{x^{1+\alpha}} \, . 
\eq
for $x \gg 1$.
Note also that because the tempering introduces the length scale, $1/\lambda$, the Green's function in this case does not have the spatio-temporal self-similar structure of Eq.~(\ref{similarity}). 

%%%%%%%%%%%%%%%%%%%%%%%%%%%%%
\subsection{Combined diffusion and fractional diffusion}
%%%%%%%%%%%%%%%%%%%%%%%%%%%%%

The last model of anomalous diffusion  that we consider combines the effects of regular diffusion and fractional diffusion. This model, which is motivated by application where the two transport mechanism are typically present, has the form
\bq
\label{frac_diff}
\partial_t \phi =\chi_d \partial^2_x \phi +\chi_f \left[ l \,
_{-\infty}D_x^\alpha +  r \, _{x}D_{\infty}^\alpha \right] \, \phi \, . 
\eq

\begin{figure}
\includegraphics[scale=0.45]{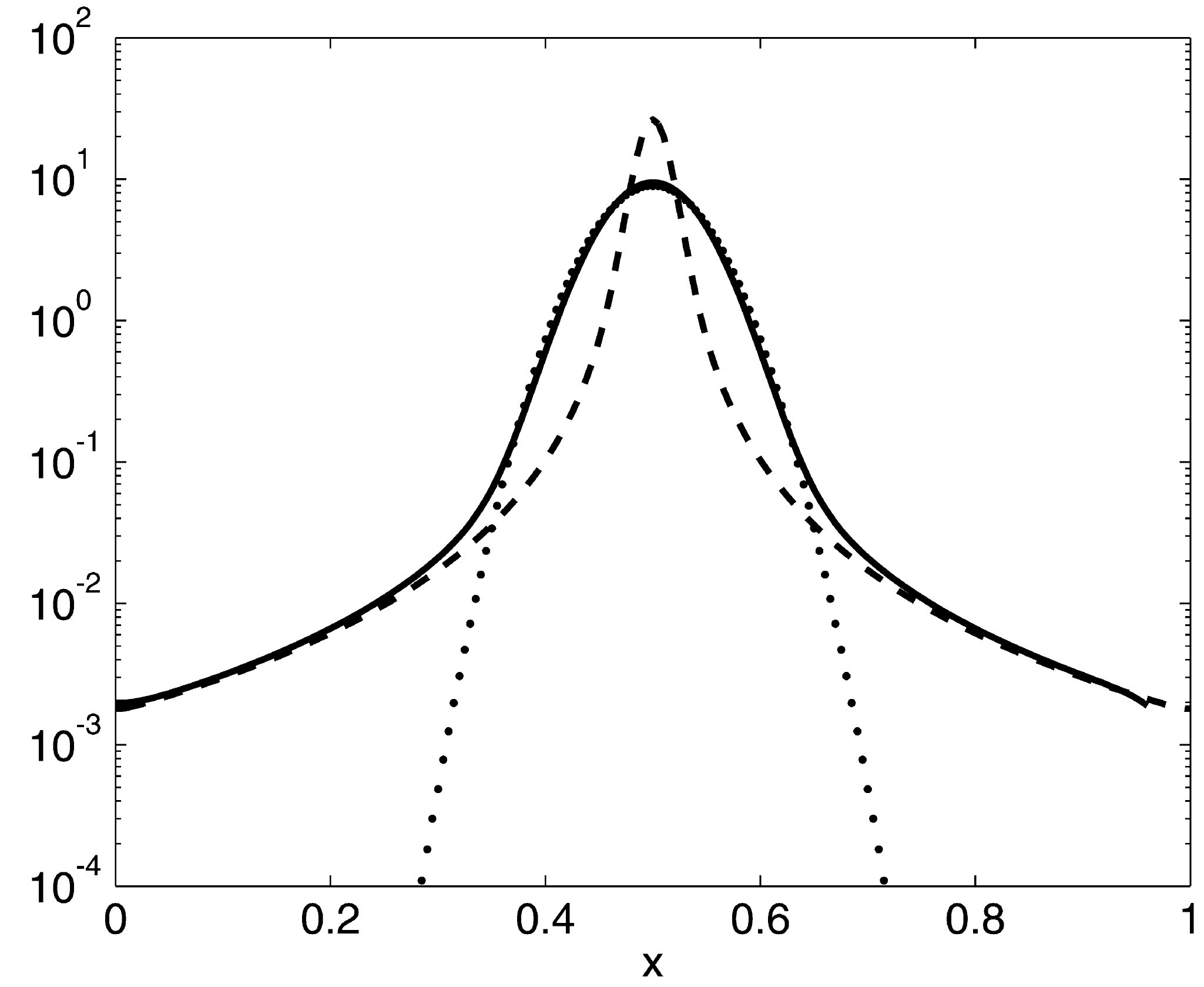}
\caption{
\label{fig_frac_diff}
Interplay of regular and fractional diffusion. 
The dotted curve corresponds to the solution of the standard diffusion Eq.~(\ref{diff}) with $\chi_d=10$, and the dashed curve corresponds
to the solution of the symmetric, $l=r$, fractional diffusion 
Eq.~(\ref{frac})  with  $\alpha=1.5$, $\theta=0$, and $\chi_f=1$.
The solid curve shows the solution of Eq.~(\ref{frac_diff}) which includes 
both, fractional and regular diffusion.}
%\vspace{10 cm}
\end{figure}

In this case the Green's function is given by Eq.~(\ref{eq_17_2}) with 
\bq
\label{psi_kk_dif}
\Lambda =-\chi_d k^2 + \chi_f \left[ l (-ik)^\alpha + r(ik)^\alpha \right]\, . % -\chi_d k^2 \, ,
\eq
Note that, like in the tempered case, in the presence of regular and fractional diffusion the Green's function is not self-similar.
As Fig.~\ref{fig_frac_diff} shows, the effects of regular diffusion are mostly present in the
core of the Green's function, and are negligible in the tails. This is
consistent with the probabilistic interpretation since regular diffusion leads to short displacements that
dominate the core of the distribution. As time advances, the Gaussian  core expands. However, for large enough $x$, the algebraic
tails persist and transport is dominated
by fractional
diffusion. In particular, the scaling in time
of the decay of the Green's function transitions from a Gaussian scaling at short times  to an asymptotic
super-diffusive scaling at long times
\bq
\label{t_scaling}
G(0,t)\sim \left\{
\begin{array}{ll}
\left ( \chi_d t \right)^{-1/2}  & {\rm for} \qquad  t \sim 0
\\ \\
\left ( \chi_f t \right)^{-1/\alpha} &  {\rm for} \qquad   t  \gg 1\, .
\end{array}
\right.
\eq

%%%%%%%%%%%%%%%%%%%%%%%%%%%%%%%%%%%%%%%%%%%%
% FIGURE
%%%%%%%%%%%%%%%%%%%%%%%%%%%%%%%%%%%%%%%%%%%%
%%%%%%%%%%%%%%%%%%%%%%%%%%%%%%%%%%%%%%%%%%%%

%%%%%%%%%%%%%%%%%%%%%%%%%%%%%%%%
\section{Front propagation}
%%%%%%%%%%%%%%%%%%%%%%%%%%%%%%%%

In this section we study the modified Fisher-Kolmogorov equation in which the 
standard diffusion operator is substituted by the anomalous transport operators  described above.
Following a brief review of basic results  on the standard Fisher-Kolmogorov equation, we study the role of fractional diffusion, tempered fractional diffusion, and combined fractional and regular diffusion in the propagation of fronts. Of particular interest is the study of the front velocity and the decay of the front's tail. 

%%%%%%%%%%%%%%%%%%%%%%%%%%%%%%%%
\subsection{Standard diffusive case}
%%%%%%%%%%%%%%%%%%%%%%%%%%%%%%%%

%%%%%%%%%%%%%%%%%%%%%%%%%%%%%%%%%%%%%%%%%%%%
% FIGURE
%%%%%%%%%%%%%%%%%%%%%%%%%%%%%%%%%%%%%%%%%%%%
\begin{figure}
\includegraphics[scale=0.4]{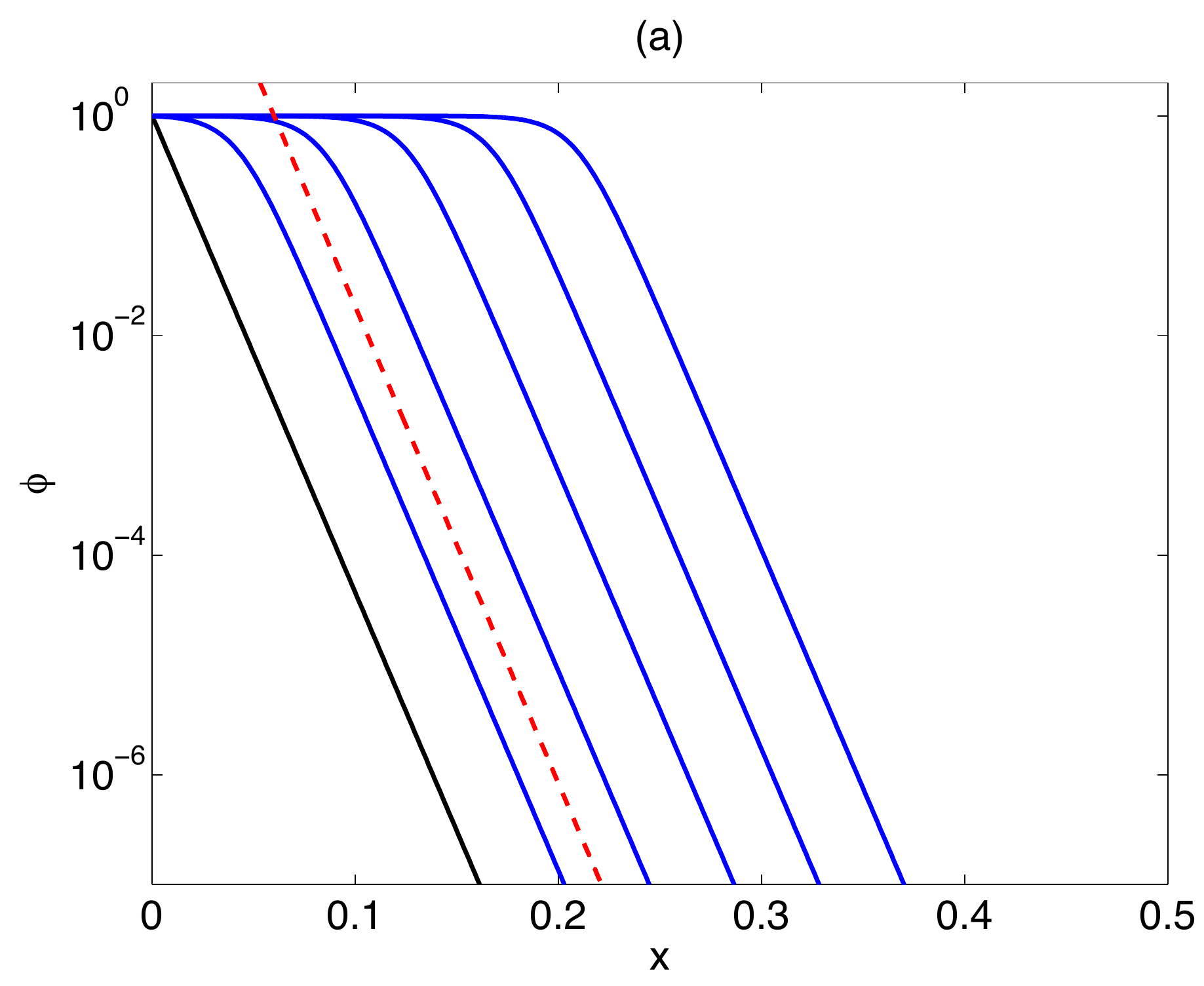}
\includegraphics[scale=0.4]{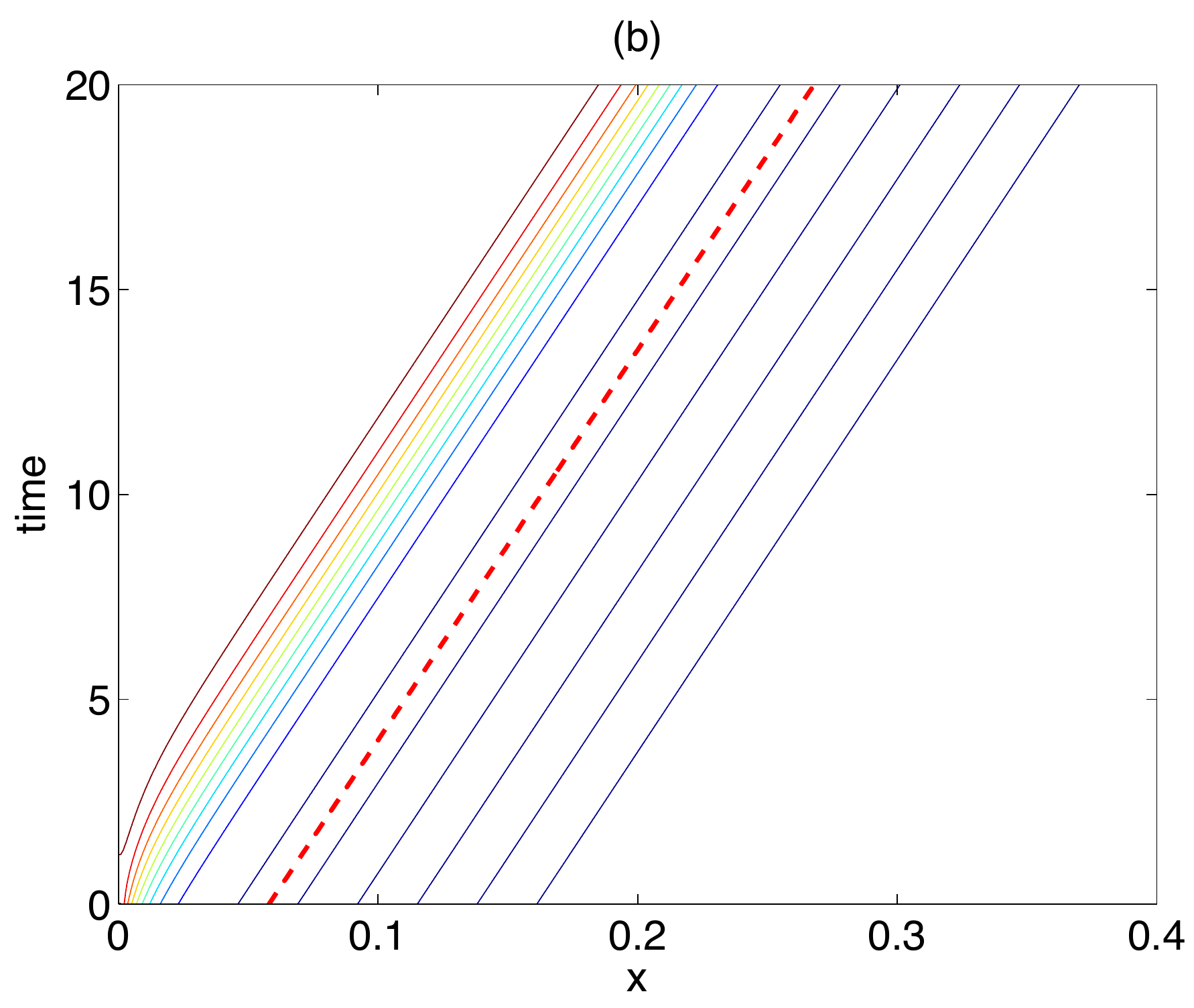}
\caption{
\label{fig_fk_diff}
Front propagation in the standard Fisher-Kolmogorov model in Eq.~(\ref{fk})
with $\chi_d=5 \times 10^{-6}$ and $\gamma=1$. Panel (a) shows the front profiles, $\phi(x,t)$, in $\log$-linear scale as functions of $x$ for different $t$. The black curve denotes the initial condition according to Eq.~(\ref{ic})
with $\nu=100$, and the blue curves show the profiles at 
$t=4,\,8,\,12,\,16,\,20$. 
The red dashed line shows the expected asymptotic decay according to Eq.~(\ref{tail_diff}). 
Panel (b) shows the isocontours $\phi(x,t)=\phi_0$ for $\phi_0=10^{-n}$ with $n=7,\,6,\,5,\,4,\,3,\,2,\,1$ (dark blue) and
$\phi_0=0.2,\, 0.3,\, 0.4 ,\,0.5 ,\,0.6 ,\,0.7,\,0.8,\,0.9,\,1$ (light blue to red).  The dashed red line 
shows the Lagrangian front speed in Eq.~(\ref{xlag_diff}).
}
%\vspace{10 cm}
\end{figure}
%%%%%%%%%%%%%%%%%%%%%%%%%%%%%%%%%%%%%%%%%%%%

For regular diffusion the model reduces to the standard Fisher-Kolmogorov equation in Eq.~(\ref{fk})
%\bq
%\label{fk_diff}
%\partial_t \phi = \chi_d \partial_x^2 \phi + \gamma \phi \left ( 1 - 
%\phi \right) \, .
%\eq
As it is well-known, in this case the reaction kinetics has two fixed points, the stable state, $\phi=1$, and the unstable state, $\phi=0$.  For initial conditions of the form
\begin{eqnarray}
\label{ic}
\phi(x,t=0)=e^{-\nu x} \, ,
%\left\{
%\begin{array}{ll}
%A
%& \mbox{ for
%}x < 0\mbox{,}
%\\ 
% B e^{-\nu x}
% & \mbox{  for } x>0
%\end{array}
%\right.
\end{eqnarray} 
%where $A$, $B$, and $\nu$ are constants.   
for $x\geq 0$, and $\phi(x,t=0)=1$ for $x<1$, the leading edge analysis implies that the tail of the  front exhibits the asymptotic
exponential decay
\bq
\label{tail_diff}
\phi \sim e^{-\nu \left( x - c t \right)}\, ,
\eq 
where the front speed, $c$, depends on the  front's  decay, $\nu$, according to
\bq
\label{c_diff}
c=\frac{\gamma}{\nu} + \nu \chi_d \, , 
%\qquad c_m=2 \sqrt{\gamma \chi}
\eq
with the minimum front velocity,
\bq
\label{c_min}
c_m=2 \sqrt{\gamma \chi_d} \, ,
\eq
corresponding to $\nu=\sqrt{\gamma/\chi_d}$,  see for example, \cite{rd_biology} and references therein.  
Figure~\ref{fig_fk_diff} shows the result of a numerical integration of  Eq.~(\ref{fk}) with 
initial condition in Eq.~(\ref{ic}). We used  an operator splitting algorithm with an explicit Euler step for the time advance of the reaction kinetics and a Crank-Nicholson semi-implicit time step for the diffusion. The diffusion operator was discretized using a centered finite-difference method. 
As Fig.~\ref{fig_fk_diff}-(a) shows, in this case the front moves rigidly with an exponential decaying tail. 
The constant propagation velocity of the front  is evident in Fig.~\ref{fig_fk_diff}-(b) that shows the spatio-temporal evolution of isocontours of $\phi$, along with the 
Lagrangian  front trajectory
\bq
\label{xlag_diff}
x_L(t)=x_{L0} + c (t - t_0) \, ,
\eq
 where $c$ is the front speed in Eq.~(\ref{c_diff}) and $t_0$ is a constant. 

%%%%%%%%%%%%%%%%%%%%%%%%%%%%%%%%
\subsection{Fractional diffusive case}
%%%%%%%%%%%%%%%%%%%%%%%%%%%%%%%%

%%%%%%%%%%%%%%%%%%%%%%%%%%%%%%%%%%%%%%%%%%%%
% FIGURE
%%%%%%%%%%%%%%%%%%%%%%%%%%%%%%%%%%%%%%%%%%%%
\begin{figure}
\includegraphics[scale=0.4]{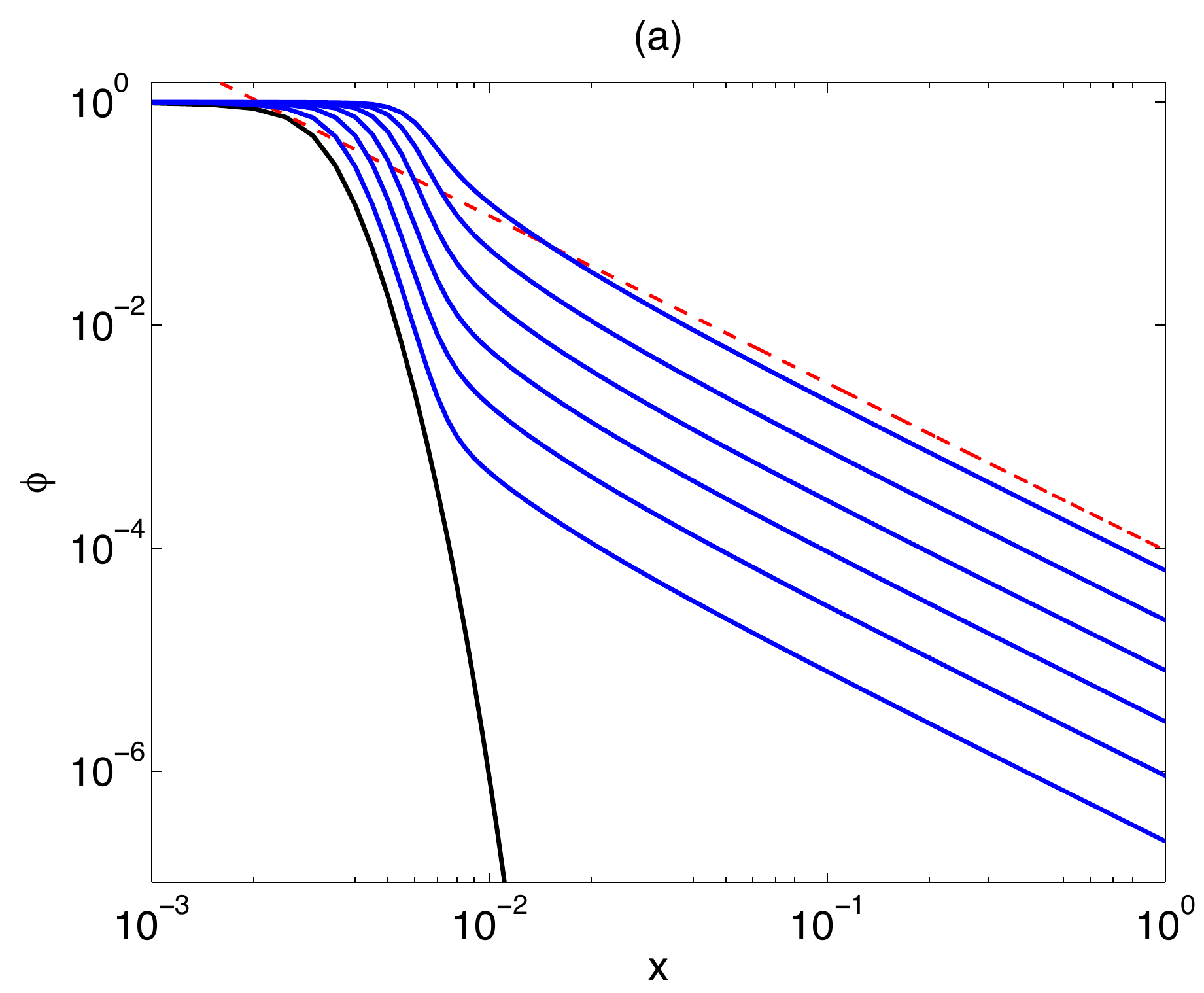}
\includegraphics[scale=0.4]{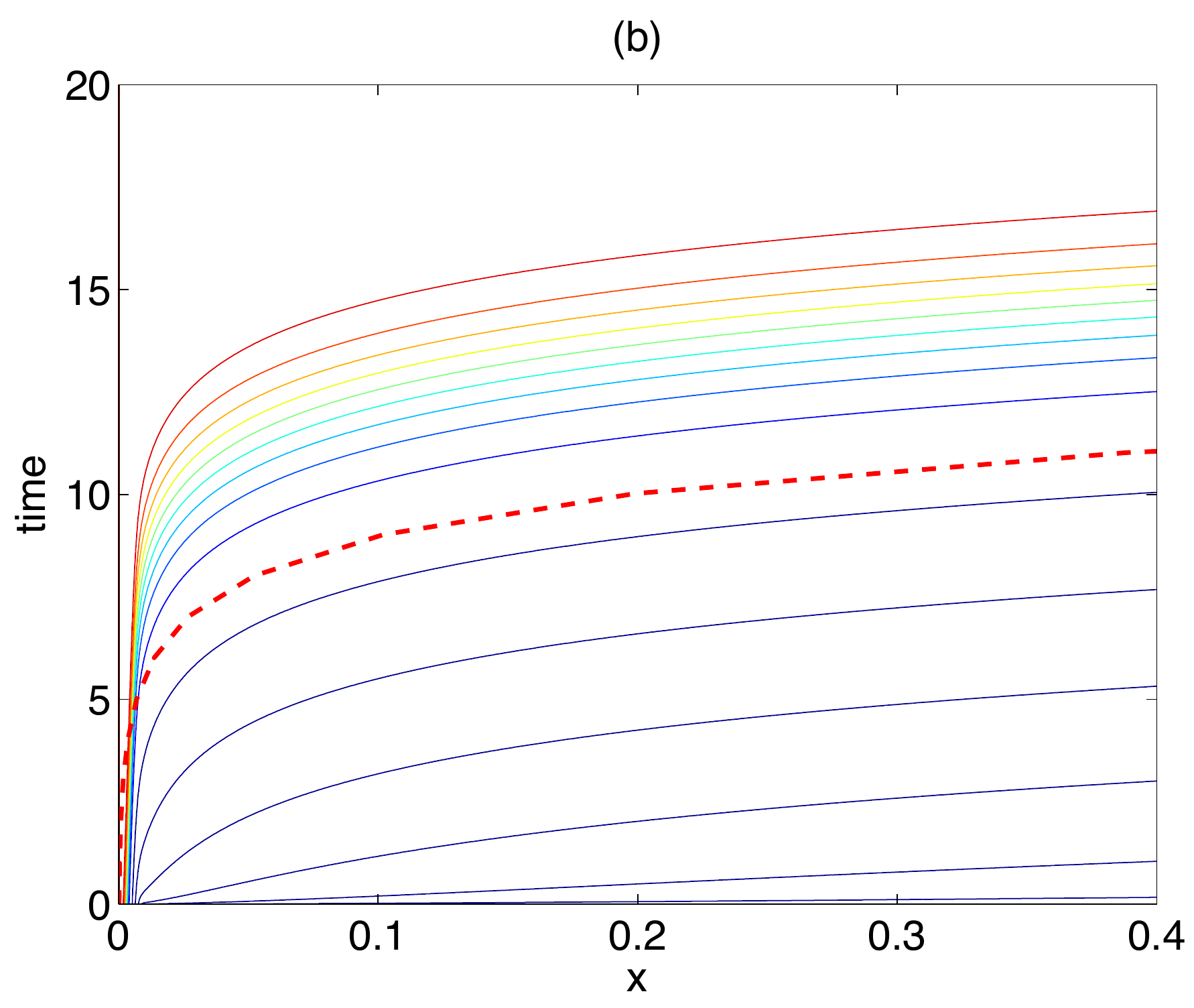}
\caption{
\label{fig_fk_frac}
Front acceleration in the fractional Fisher-Kolmogorov model in Eq.~(\ref{fk_frac})
with $\alpha=1.5$,  $\chi_f=5 \times 10^{-7}$ and $\gamma=1$. 
Panel (a) shows the front profiles $\phi(x,t)$ in $\log$-$\log$ scale as functions of $x$ for different $t$. The black curve denotes the initial condition,  and the blue curves show the profiles at 
$t=1.05,\, 2.10,\, 3.15,\, 4.20,\, 5.25,\, 6.30$. 
The red dashed line shows the expected asymptotic decay according to Eq.~(\ref{tail_frac_1}). 
Panel (b) shows the isocontours $\phi(x,t)=\phi_0$ for $\phi_0=10^{-n}$ with $n=7,\,6,\,5,\,4,\,3,\,2,\,1$ (dark blue) and
$\phi_0=0.2,\, 0.3,\, 0.4 ,\,0.5 ,\,0.6 ,\,0.7,\,0.8,\,0.9,\,1$ (light blue to red).  The dashed red line 
shows the asymptotic Lagrangian front path according to Eq.~(\ref{lagx}).
}
%\vspace{10 cm}
\end{figure}
%%%%%%%%%%%%%%%%%%%%%%%%%%%%%%%%%%%%%%%%%%%%

To study the effect of fractional diffusion in the propagation of fronts we consider 
the left factional Fisher-Kolmogorov equation
\bq
\label{fk_frac}
\partial_t \phi = \chi_f\, _{-\infty}D_x^\alpha \phi + \gamma \phi \left ( 1 - 
\phi \right) \, ,
\eq
where $_{-\infty}D_x^\alpha$ is the left Riemann-Liouville fractional derivative in Eq.~(\ref{left}). 
This model was originally introduce in Ref.~\cite{del_castillo_2003} to study the interplay of the reaction kinetics with anomalous diffusion processes that exhibit L\'evy flights in one direction, say, for $x > 0$, but Gaussian behavior in the other direction, $x < 0$. 

The leading edge analysis of Eq.~(\ref{fk_frac}) reveals two unique features of front propagation in the presence of fractional diffusion: (i) front acceleration and (ii) algebraic decaying of the front's tail 
\cite{del_castillo_2003}. 
In particular, for a large, fixed $t$, in the limit $x \left( \chi_f t \right)^{-1/\alpha} \rightarrow \infty$, 
\bq
\label{tail_frac_1}
%% 
 % \phi \sim\chi t e^{\gamma t} \left[ \frac{A}{\alpha} x^{-\alpha} + 
 % \frac{1}{\nu} x^{-1-\alpha} + {\cal C} e^{-\nu x}\right] 
 % \, .
 %%
\phi \sim %\chi t e^{\gamma t} \left[ 
\frac{1}{\alpha} x^{-\alpha} + 
\frac{1}{\nu} x^{-1-\alpha} + \dots %\right] 
\, . 
\eq
On the other hand, the time-asymptotic dynamics for fixed, large $x$ gives to leading order $\phi \sim e^{\gamma t}$ with correction of the order $\left( \chi_f t\right)^{-1/\alpha} e^{\gamma t}$. These results imply the spatio-temporal asymptotic scaling 
\bq
\label{xt_asym}
\phi \sim x^{-\alpha} e^{\gamma t} \, .
\eq
For $\phi_L \in (0,1)$, the solution of the equation $\phi( x,t)=\phi_L$, 
gives the Lagrangian trajectory, $x_L(t)=x(t; \phi_L)$, of the coordinate, $x_L$,  of a point in the front with concentration $\phi_L$. In the case when $\phi$ satisfies the scaling in Eq.~(\ref{xt_asym}), 
%From Eq.~(\ref{xt_asym}) it then follows, 
\bq
\label{lagx}
%x_L(t)= \left( \frac{\chi_f t }{\alpha}\right)^{1/\alpha} \, e^{\gamma t / \alpha} \, ,
%x_L(t)=x_0+ \phi_0 \left( \chi_f t \right)^{1/\alpha} \, e^{\gamma t / \alpha} \, .
x_L(t)= x_{L0} e^{\gamma (t-t_0)/\alpha} \, ,\eq
where $\phi_L=\phi(x_{L0},t_0)$. Equation~(\ref{lagx})  
implies an exponential Lagrangian acceleration of the front, 
$a_L(t) \sim e^{\gamma t/\alpha}$. 

For the numerical integration of Eq.~(\ref{fk_frac}) in the finite-size domain $x\in (0,1)$,
the unbounded fractional derivative, $ _{-\infty}D_x^\alpha$, needs to be truncated. 
However, care must be taken doing this because the left Riemann-Liouville derivative, 
$_aD_x^\alpha \phi$, with $a$ finite is in general singular at the lower boundary $x=a$. 
To circumvent this problem, following Ref.~\cite{del_castillo_2003}, we use the regularized (in the Caputo sense) fractional derivative of the form 
$_a^CD_x^\alpha \phi=_aD_x^\alpha \left[ \phi(x)-\phi(a)-\phi'(a)(x-a) \right]$, which in this case reduces to 
$_0D_x^\alpha \left[ \phi(x)-1 \right]$ due to the boundary conditions at $x=0$. For a  detailed discussion of the regularization of general fractional diffusion models in finite-size domains, see 
Ref.~\cite{del_castillo_2006}. 
Like in the standard Fisher-Kolmogorov model, for the numerical integration of Eq.~(\ref{fk_frac})
we use  an operator splitting algorithm with an explicit Euler step for the time advance of the reaction kinetics and a Crank-Nicholson semi-implicit time step for the fractional diffusion. 
However, in the fractional case, we use a flux-conserving 
scheme with an upwind Grunwald-Letnikov finite-difference discretization of the regularized fractional derivative. Further details of the numerical  method can be found in Ref.~\cite{del_castillo_2006}. 
Figure~\ref{fig_fk_frac} shows the results of the numerical integration of 
the fractional Fisher-Kolmogorov model in Eq.~(\ref{fk_frac}) for an initial condition of the form
$\phi(x,t=0)=\left[ 1 - \tanh \left((x-x_0)/W\right) \right]/2$ with $x_0=0.003$, $W=0.001$. 
In agreement with the leading edge asymptotic result in Eq.~(\ref{tail_frac_1}), it is observed that the front develops and algebraic decaying tail.  Clear evidence of the acceleration of the front is shown in the spatio-temporal evolution of the front's isocontours $\phi(x,t)=\phi_0$ shown in panel (b). 

%%%%%%%%%%%%%%%%%%%%%%%%%%%%%%%%
\subsection{Tempered fractional diffusive case}
%%%%%%%%%%%%%%%%%%%%%%%%%%%%%%%%

%%%%%%%%%%%%%%%%%%%%%%%%%%%%%%%%%%%%%%%%%%%%
% FIGURE
%%%%%%%%%%%%%%%%%%%%%%%%%%%%%%%%%%%%%%%%%%%%
\begin{figure}
\includegraphics[scale=0.4]{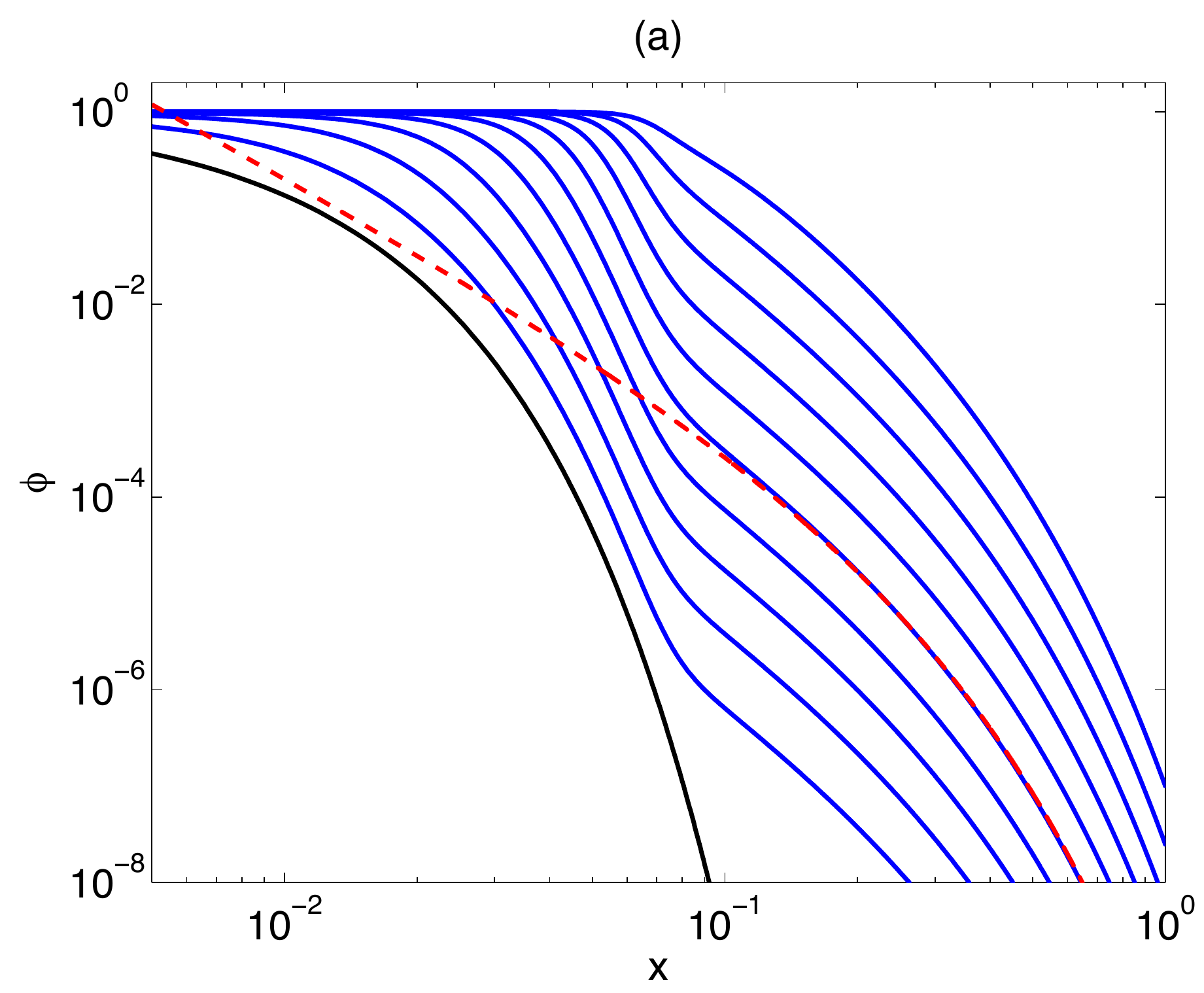}
\includegraphics[scale=0.4]{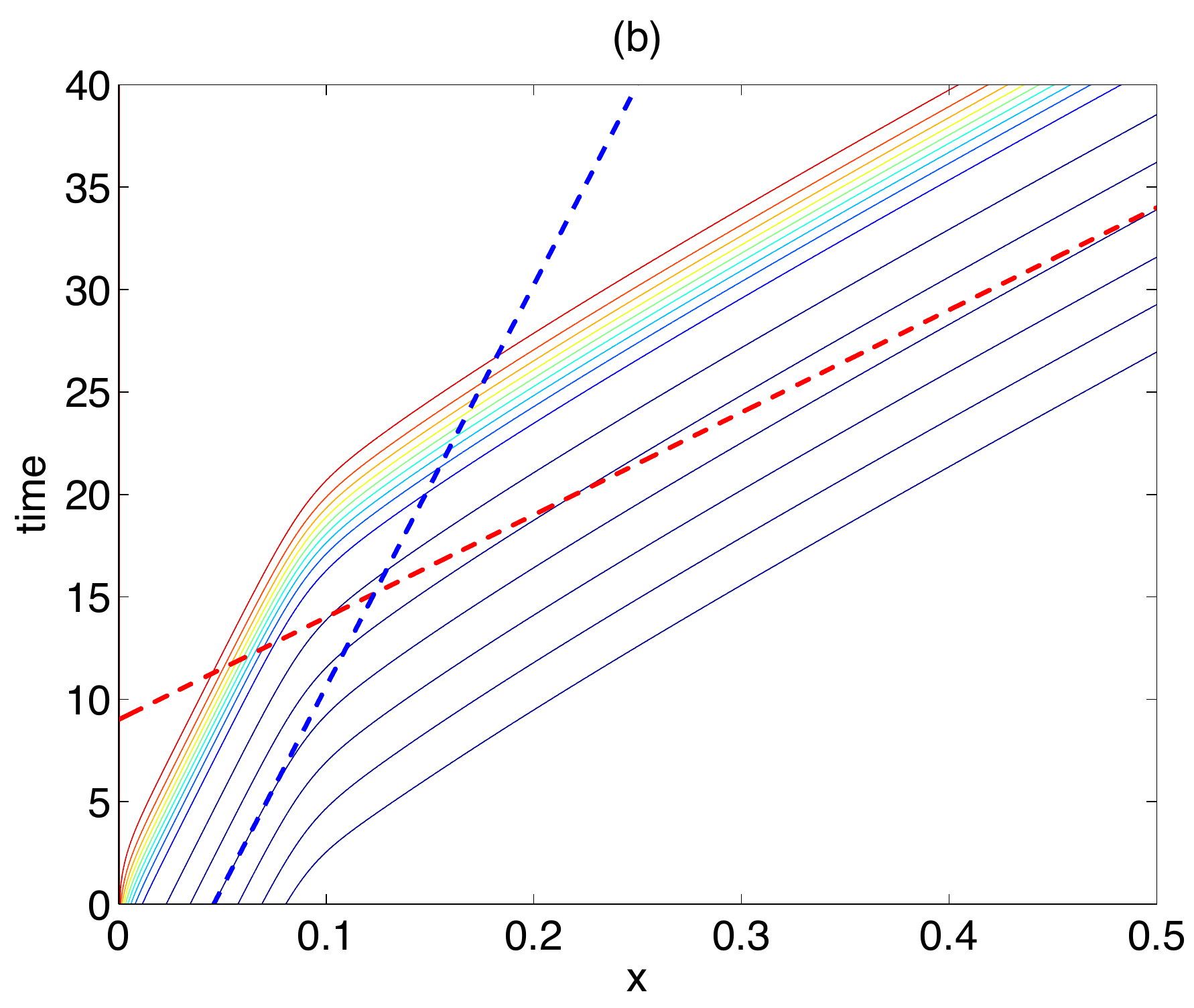}
\caption{
\label{fig_fk_temp}
Front propagation in the tempered fractional Fisher-Kolmogorov model in Eq.~(\ref{tempered_fk})
with $\alpha=1.5$,  $\lambda=50$,  $\chi_\tau=5 \times 10^{-7}$ and $\gamma=1$. 
Panel (a) shows the front profiles $\phi(x,t)$ in $\log$-$\log$ scale as functions of $x$ for different $t$. The black curve denotes the initial condition in Eq.~(\ref{ic})
with $\nu=200$, and the blue curves show the profiles at 
$t=1.4,\, 2.8,\, 4.2,\, 5.6,\, 7,\, 8.4,\, 9.8,\, 11.2,\, 12.6,\,14$.
The red dashed line shows the expected asymptotic decay according to Eq.~(\ref{tail_temp_1}). 
Panel (b) shows the isocontours $\phi(x,t)=\phi_0$ for $\phi_0=10^{-n}$ with $n=7,\,6,\,5,\,4,\,3,\,2,\,1$ (dark blue) and
$\phi_0=0.2,\, 0.3,\, 0.4 ,\,0.5 ,\,0.6 ,\,0.7,\,0.8,\,0.9,\,1$ (light blue to red).  The 
dashed blue line shows the transient 
Lagrangian front path $x=x_0+c t$ where $c$ is the diffusive front speed in Eq.~(\ref{c_diff})
and $x_0$ a constant. 
The dashed red line shows the asymptotic Lagrangian front path $x=x_*+v_* t$ where $v_*$ is the terminal velocity in Eq.~(\ref{lagv_temp_3}) and $x_*$ a constant.
}
%\vspace{10 cm}
\end{figure}
%%%%%%%%%%%%%%%%%%%%%%%%%%%%%%%%%%%%%%%%%%%%

%%%%%%%%%%%%%%%%%%%%%%%%%%%%%%%%%%%%%%%%%%%%
% FIGURE
%%%%%%%%%%%%%%%%%%%%%%%%%%%%%%%%%%%%%%%%%%%%
\begin{figure}
\includegraphics[scale=0.4]{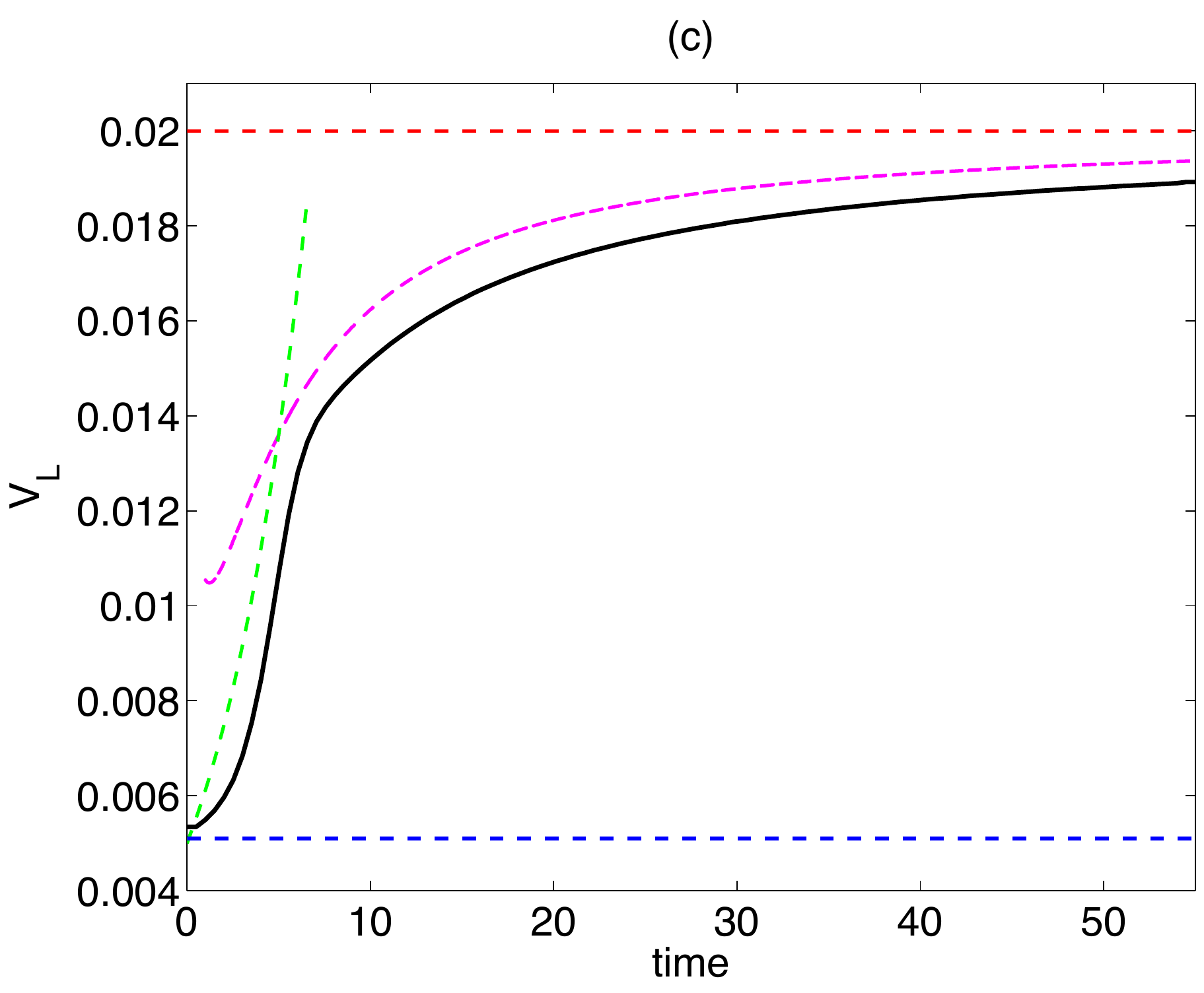}
\caption{
\label{fig_fk_vel_temp}
Lagrangian front velocity as function of time  in the tempered fractional Fisher-Kolmogorov model in Eq.~(\ref{tempered_fk}) with $\alpha=1.5$,  $\lambda=50$,  $\chi_\tau=5 \times 10^{-7}$ and $\gamma=1$. 
The solid black line denotes the numerical result. The horizontal blue (red) dashed line at the bottom (top) denotes $c$ in Eq.~(\ref{c_diff}) ($v_*$ in Eq.~(\ref{lagv_temp_3})). 
The green dashed line denotes the 
transient exponential front velocity according to Eq.~(\ref{lagx}).The magenta dashed line denotes the intermediate asymptotic front velocity in Eq.~(\ref{lagv_temp_1}). 
}
%\vspace{10 cm}
\end{figure}
%%%%%%%%%%%%%%%%%%%%%%%%%%%%%%%%%%%%%%%%%%%%

To study the role of truncation in the super-diffusive  acceleration 
of fronts due to L\'evy flights, we consider 
the fractional Fisher-Kolmogorov Eq.~(\ref{fk_frac}) and substitute the left Riemann-Liouville 
fractional derivatives by the left truncated fractional derivative in 
Eq.~(\ref{temp_1}),
\bq
\label{tempered_fk}
\partial_t \phi = \chi_\tau \left[ e^{-\lambda x}\,
_{-\infty}D_x^\alpha\, \left( e^{\lambda x} \phi \right) - \lambda^{\alpha}\phi \right] 
+ \gamma \phi \left( 1 - \phi \right )
\, ,
\eq
where $\chi_\tau= \chi_t/ \left| \cos \alpha \pi/2 \right |$. 
Note that, to simplify the analysis, we have assumed that there is no drift velocity, i.e., $V=0$. 
According to the discussion following Eq.~(\ref{temp_1}), for the $1<\alpha <2$ case of interest here, this assumption implies that 
$V=V_D-v=0$.  That is, it is assumed that there is a background drift, $V_D$, that cancels the drift, $v$,  resulting from the bias of the stochastic process.

There are two characteristic time scales in this problem. 
One is the crossover time scale, $\chi_\tau t_c= \lambda^{-\alpha}$, and the other is  
the reaction time scale,
\bq
t_r=1/\gamma \, .
\eq
Note also that due to the term $\chi_\tau \lambda^\alpha$, the tempering of the fractional Fisher-Kolmogorov equation introduces an ``effective" reaction rate, $\gamma_{eff}=\gamma-\chi_\tau \lambda^\alpha$.  Here we limit attention to the case $\gamma_{eff}>0$, i.e. we assume that the cross-over time scale is longer than the reaction time scale, $t_c>t_r$. 

Figure~\ref{fig_fk_temp} shows the results of the numerical  integration of the 
the tempered fractional Fisher-Kolmogorov model in Eq.~(\ref{tempered_fk})
with $\alpha=1.5$,  $\lambda=50$,  $\chi_\tau=5 \times 10^{-7}$ and $\gamma=1$, for the initial condition in Eq.~(\ref{ic}) with $\nu=200$ for $x>0$, and $\phi_0=0$ for $x<0$. 
Because of the tempering, the front does not exhibit the algebraic decaying tail observed in the 
fractional-diffusion case. However, it does not decay exponentially like in the standard diffusive case either. In fact, 
if $\nu>\lambda>0$, i.e., when the initial condition decays faster than 
the truncation, the leading edge analysis of Eq.~(\ref{tempered_fk}) leads to the asymptotic scaling \cite{del_castillo_2009}
\bq
\label{tail_temp_1}
\phi \sim \left[ \frac{\chi_\tau \, t}{\nu - \lambda}
%+ \frac{A}{\lambda}
\right ]\, 
\frac{ e^{-\lambda x}}{x^{1+\alpha}}\, e^{\left( \gamma - \chi_\tau 
\lambda^\alpha \right) t} \, ,
\eq
for $x \gg \left( \chi_\tau t\right)^{1/\alpha}$. In particular, for fixed $t$, as shown in 
Fig.~\ref{fig_fk_temp}-(a), the spatial decay of the front's tail exhibits the tempered decay in 
Eq.~(\ref{temp_decay}).
Figure~\ref{fig_fk_temp}-(b) shows the typical three stages in the dynamics of tempered fronts:
(i) transient constant velocity propagation; (ii) front acceleration in the intermediate asymptotic regime;  (iii) asymptotic  approach to terminal constant propagation velocity. 
In the specific case shown here, for $\phi=10^{-4}$, the transient period spans  $0<t \lesssim 6$, and during that time the front propagates at the diffusive speed in Eq.~(\ref{c_diff}). 
During $6 \lesssim t \lesssim 30$ the front exhibits an acceleration phase that eventually leads to a state with a constant terminal propagation velocity  as $t \rightarrow \infty$ \cite{del_castillo_2009}. 

As in the case of the fractional fronts, we define the Lagrangian trajectory of the front as the function $x_L=x_L(t)$ such that $\phi(x_L,t)=\phi_L$ where $\phi_L$ is a constant.
From Eq.~(\ref{tail_temp_1}) it then follows that,  for $x>1/\lambda$,
\bq
\label{lagx_temp}
-\lambda x_L(t) + \left( \gamma - \chi_\tau \lambda^\alpha\right) t + \ln t 
-(\alpha+1) \ln x_L(t) = M \, ,
\eq
where $M$ is a constant that depends on $\phi_L$. Taking the time derivative of 
Eq.~(\ref{lagx_temp}) gives the Lagrangian speed of the front,
\bq
\label{lagv_temp_1}
v_L(t)=\frac{d x_L}{dt}=\frac{\gamma - \chi_\tau 
\lambda^\alpha+\frac{1}{t}}{\lambda + \frac{\alpha+1}{x_L(t)}}\, ,
\eq
which in the limit $t \rightarrow \infty$, gives the terminal front velocity, $v_L(t) \rightarrow v_*$, 
\bq
\label{lagv_temp_3}
v_*= \frac{\gamma-\lambda^\alpha \chi_\tau}{\lambda}\, .
\eq
The asymptotic approach to the terminal velocity is clearly observed in Figure~\ref{fig_fk_temp}-(b), and in Fig.~\ref{fig_fk_vel_temp} that shows the detailed temporal evolution of the Lagrangian velocity. 
It is interesting to observe that the decay of  the Lagrangian front acceleration, $a_L=d v_L(t)/dt$, 
\bq
\label{laga_temp_1}
a_L(t)=\frac{v_L(t)}{t\left( \lambda t v_*+1\right)}
\left\{ \left( \alpha+1\right) \left[\frac{v_L(t) 
}{x_L(t)/t}\right]^2-1\right\}
\, ,
\eq
exhibits to leading order the algebraic scaling,
\bq
\label{laga_temp_2}
a_L(t) \sim  \frac{\alpha}{\lambda t^2} \, ,
\eq
at $t \rightarrow \infty$. 

%%%%%%%%%%%%%%%%%%%%%%%%%%%%%%%%
\subsection{Fractional fronts with diffusion case}
%%%%%%%%%%%%%%%%%%%%%%%%%%%%%%%%

\begin{figure}
\includegraphics[scale=0.3]{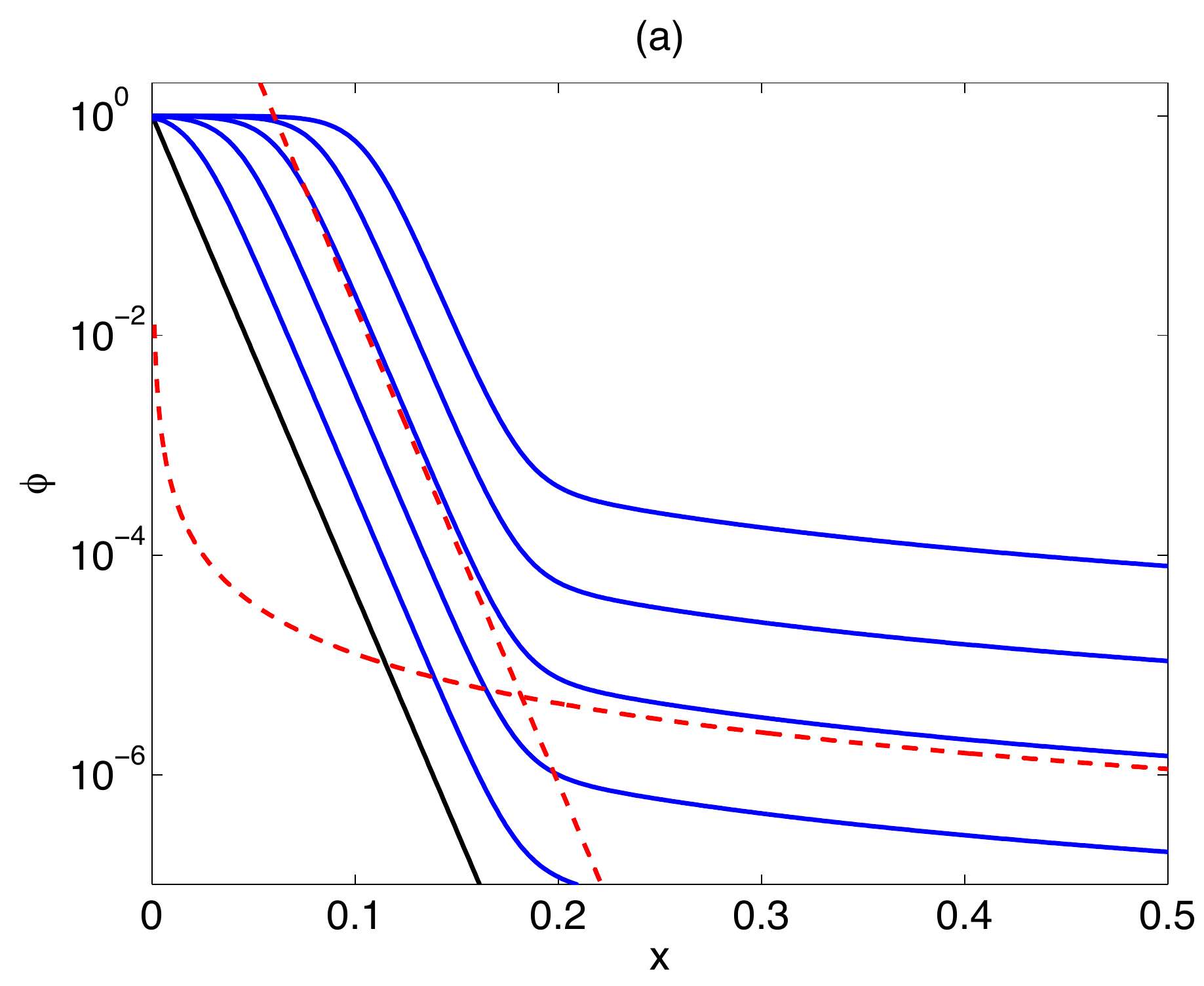}
\includegraphics[scale=0.3]{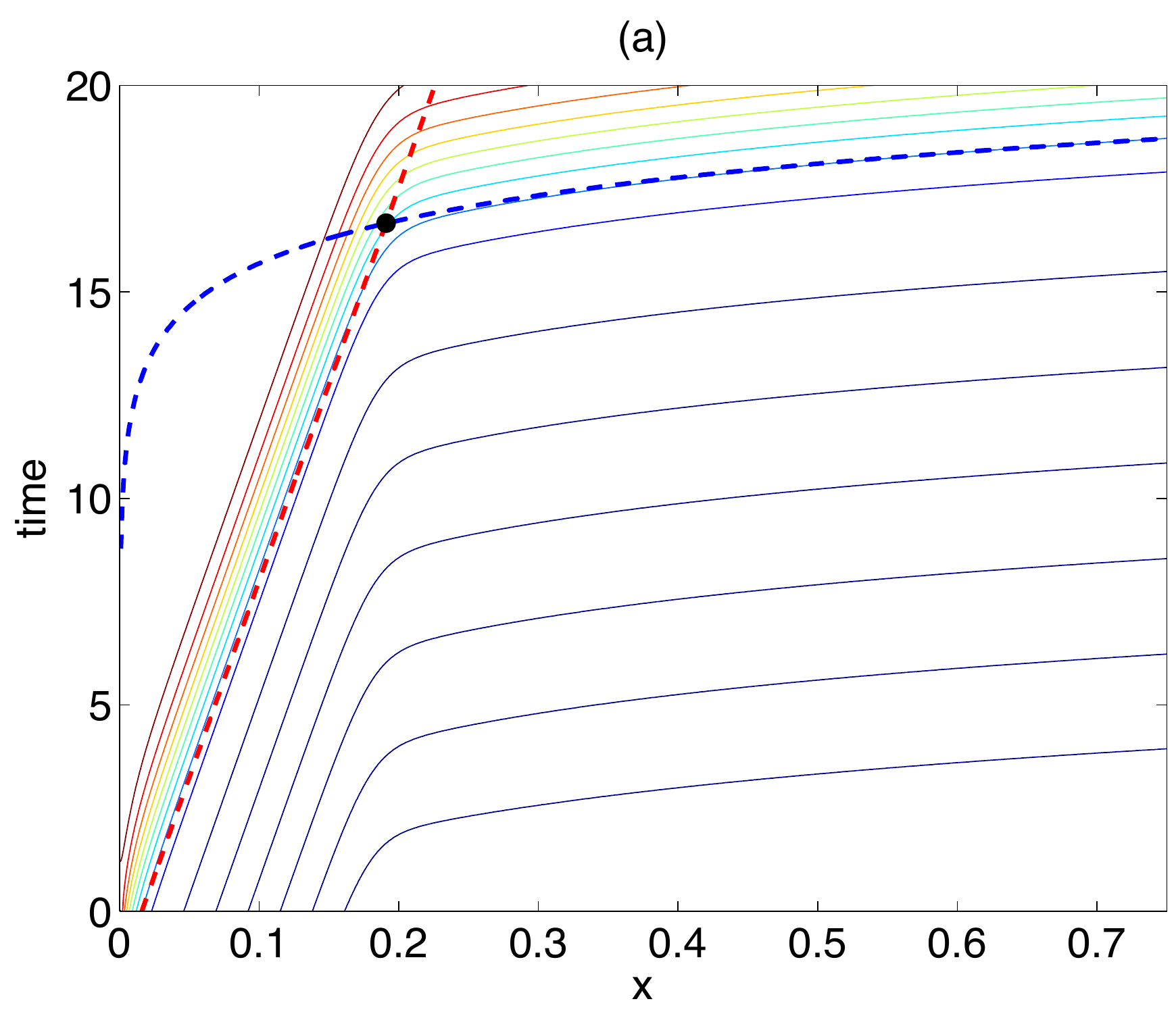}
\includegraphics[scale=0.3]{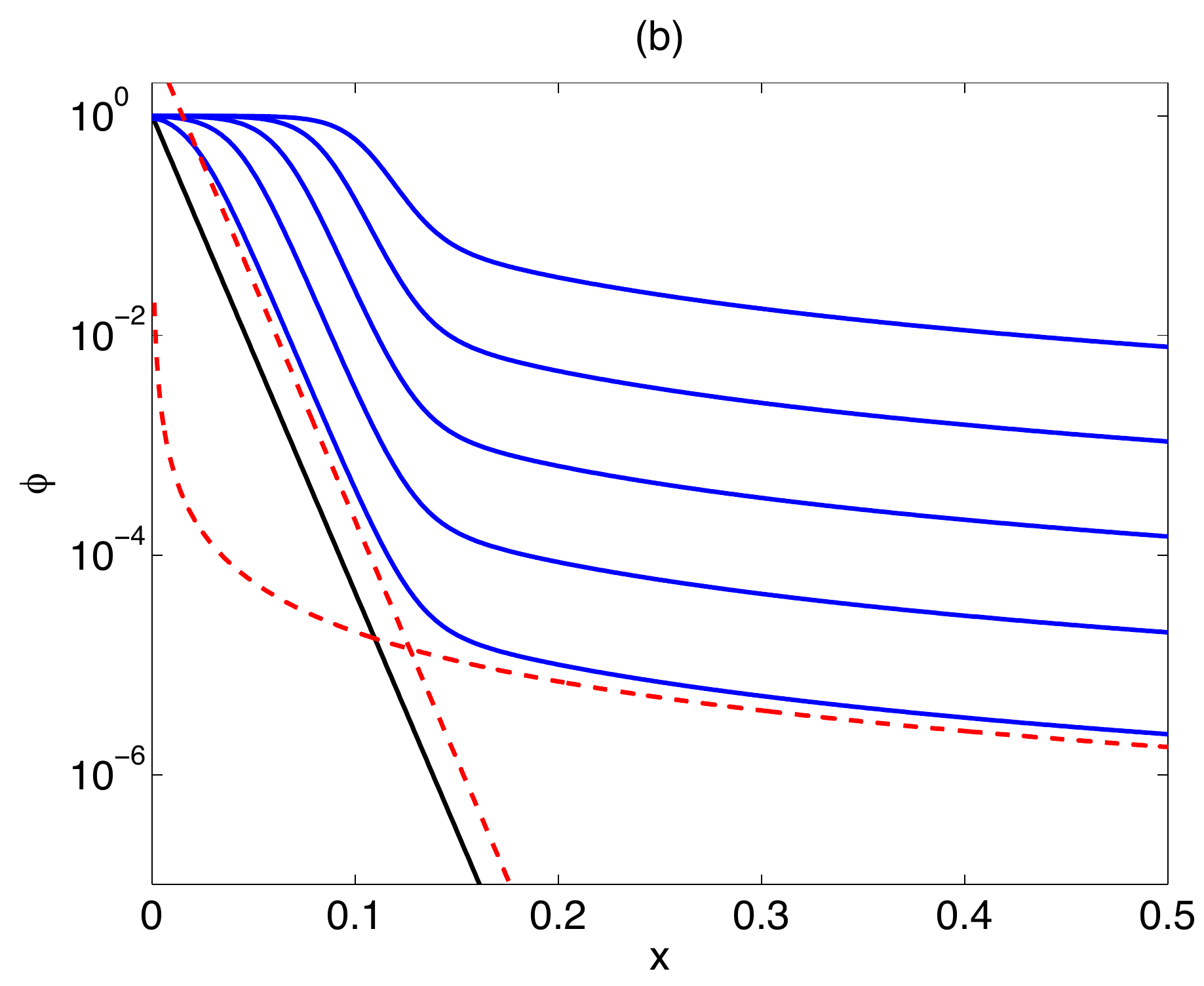}
\includegraphics[scale=0.3]{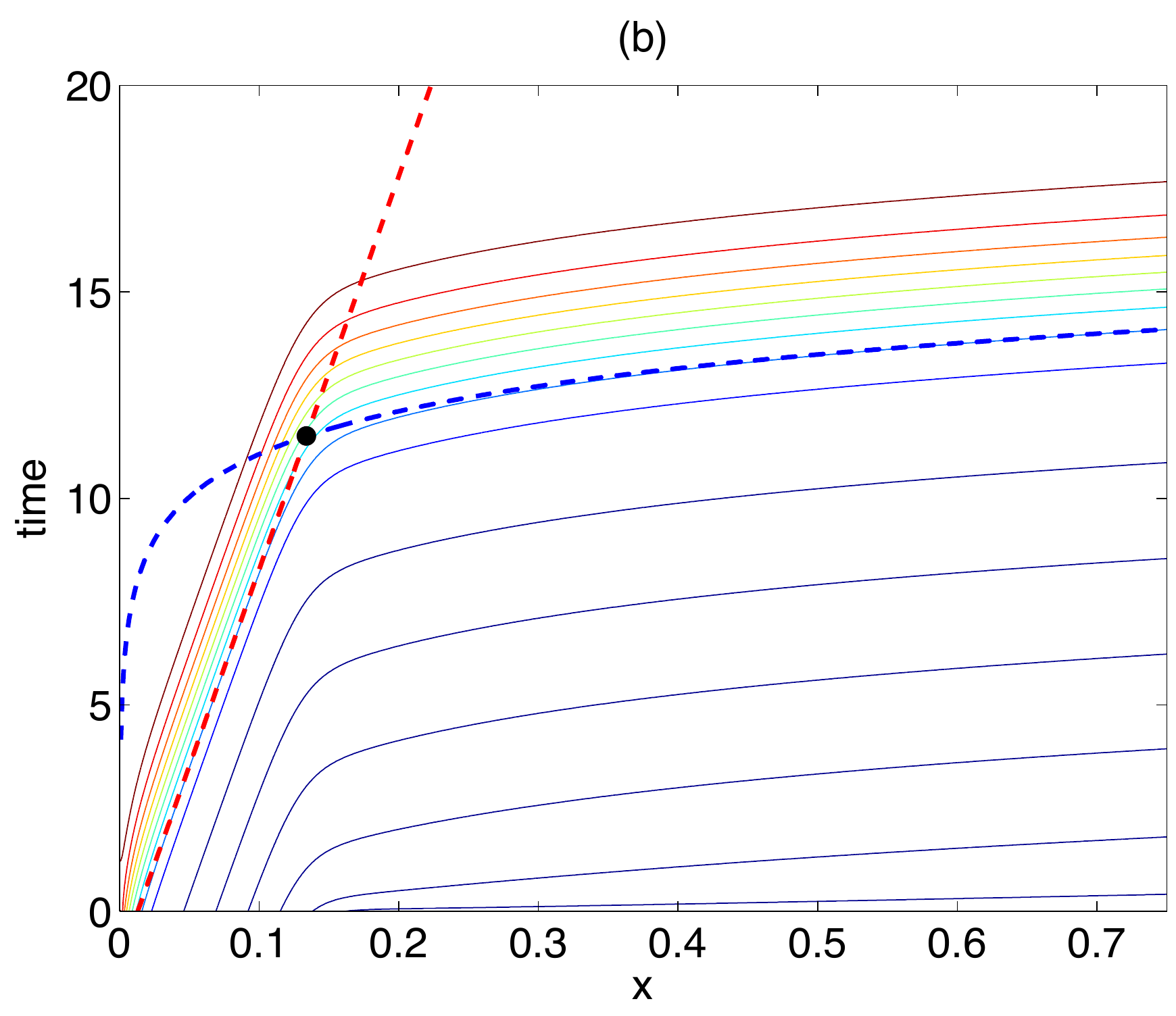}
\includegraphics[scale=0.3]{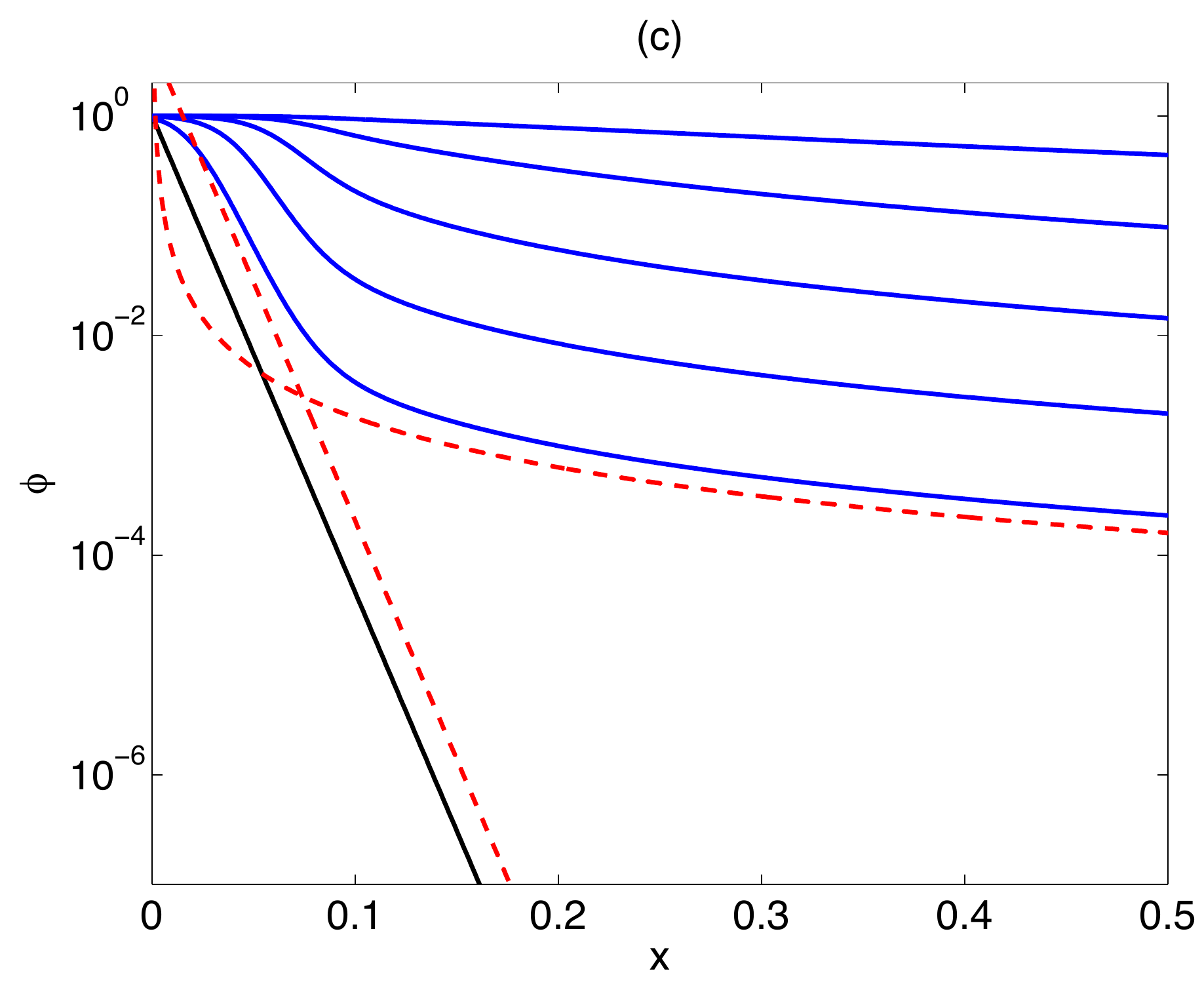}
\includegraphics[scale=0.3]{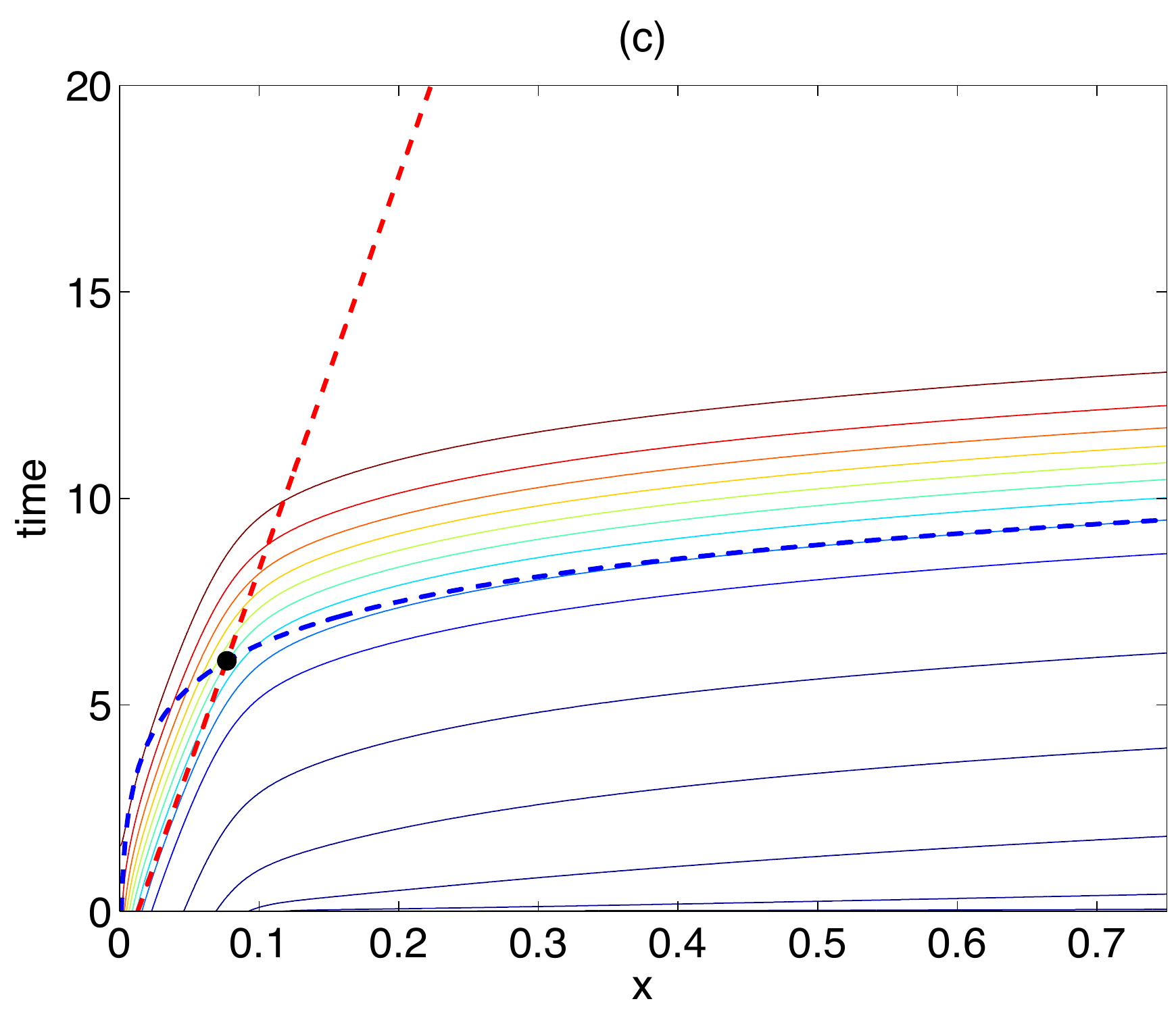}
\caption{
\label{fig_fk_frac_diff}
Front propagation in the fractional Fisher-Kolmogorov model in the presence of diffusion
according to Eq.~(\ref{fk_frac_diff}) with  $\alpha=1.5$,  $\chi_d=5 \times 10^{-6}$ and $\gamma=1$.
Cases (a), (b) and (c) correspond to  $\chi_f=5 \times 10^{-9}$,  $\chi_f=5 \times 10^{-7}$,
and  $\chi_f=5 \times 10^{-5}$ respectively. The left column shows  
the front profiles $\phi(x,t)$ in $\log$-lin scale as functions of $x$ for different $t$. The black curve denotes the initial condition in Eq.~(\ref{ic}) with $\nu=100$,
and the blue curves show the profiles at 
$t=2,\, 4,\, 6,\, 8,\, 10$.
The red dashed straight line shows the  short time decay according to Eq.~(\ref{tail_diff}). 
The red dashed curve shows the  asymptotic fractional decay according to Eq.~(\ref{tail_frac_1}). 
The right column shows
the isocontours $\phi(x,t)=\phi_0$ for $\phi_0=10^{-n}$ with $n=7,\,6,\,5,\,4,\,3,\,2,\,1$ (dark blue) and
$\phi_0=0.2,\, 0.3,\, 0.4 ,\,0.5 ,\,0.6 ,\,0.7,\,0.8,\,0.9,\,1$ (light blue to red).
The  intersection of the Lagrangian diffusive front path in Eq.~(\ref{xlag_diff}) (red dashed line)
with the  Lagrangian fractional diffusion front path in Eq.~(\ref{lagx}) (blue dashed line) gives the spatio-temporal cross-over scale $(x_c,t_c)$. 
}
%\vspace{10 cm}
\end{figure}

As mentioned before, in most applications it is natural to expect that anomalous diffusion processes will act 
in conjunction with standard diffusion. It is thus of significant interest to study the interplay of fractional dynamics and standard diffusion.  To explore this problem in the context of  reaction-diffusion fronts, we consider 
\bq
\label{fk_frac_diff}
\partial_t \phi =\chi_d \partial_x^2 \phi + \chi_f\,  _{-\infty}D_x^\alpha \phi + \gamma \phi \left ( 1 - 
\phi \right) \, .
\eq
Figure~\ref{fig_fk_frac_diff} summarize the results of the numerical integration of Eq.~(\ref{fk_frac_diff}) 
with $\alpha$, $\gamma$, and $\chi_d$ fixed, and different values of the fractional diffusivity, $\chi_f$. 
The plots on the left column show that the spatial dependence of the profiles of the fronts exhibit 
a transition from exponential decay (characteristic of diffusive fronts) to 
algebraic  decay (characteristic of fractional fronts) with the crossover distance inversely proportional to 
$\chi_f$. On the other hand, the spatio-temporal plots on the right column of Fig.~\ref{fig_fk_frac_diff} indicate  that  standard diffusion leads to a time delay in the onset of the front acceleration. 
In particular, the smaller  the ratio, $\chi_f/\chi_d$, the longer the transient phase during which the front propagates at a  constant speed and the longer it takes for the front acceleration to kick in. 
 
\begin{figure}
\includegraphics[scale=0.4]{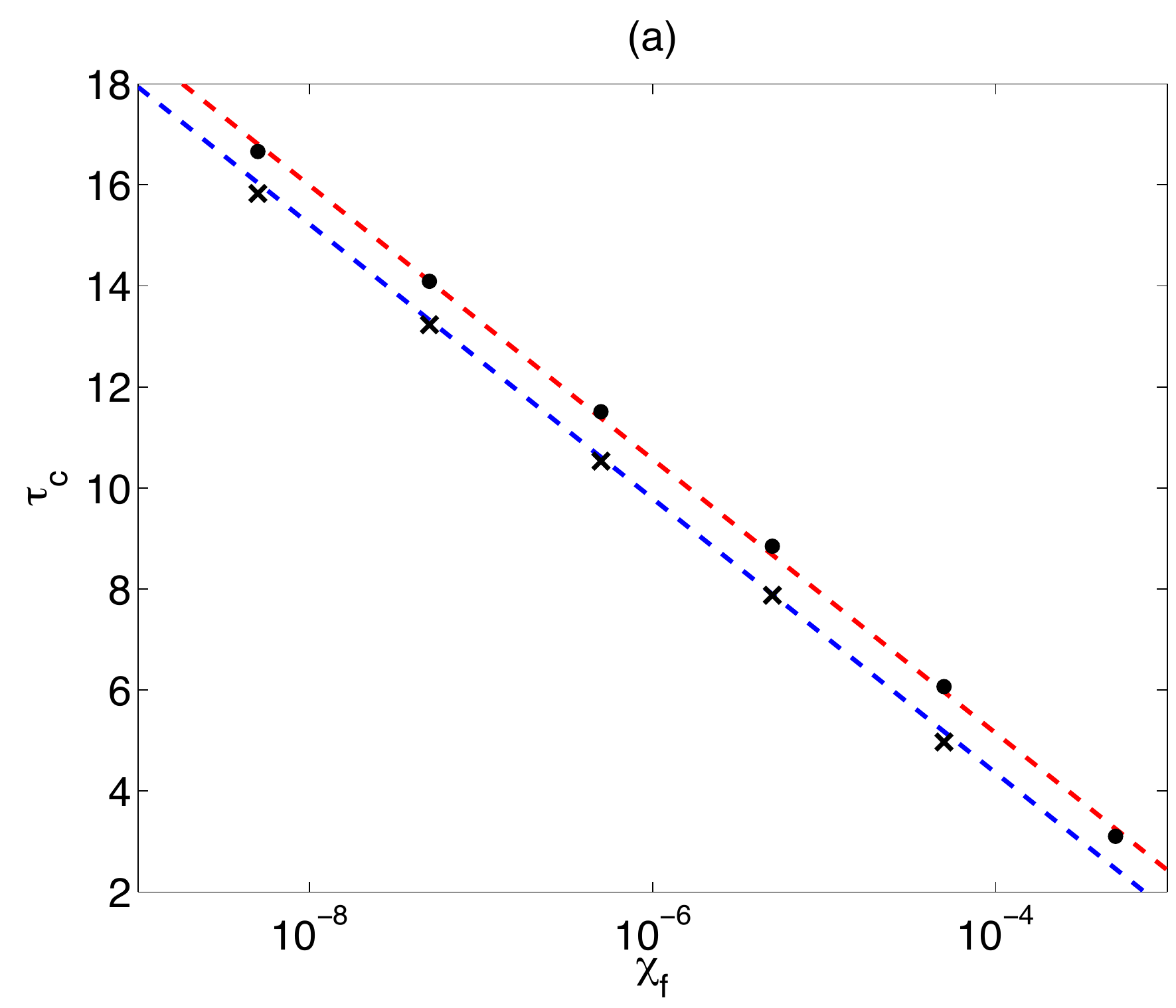}
\includegraphics[scale=0.4]{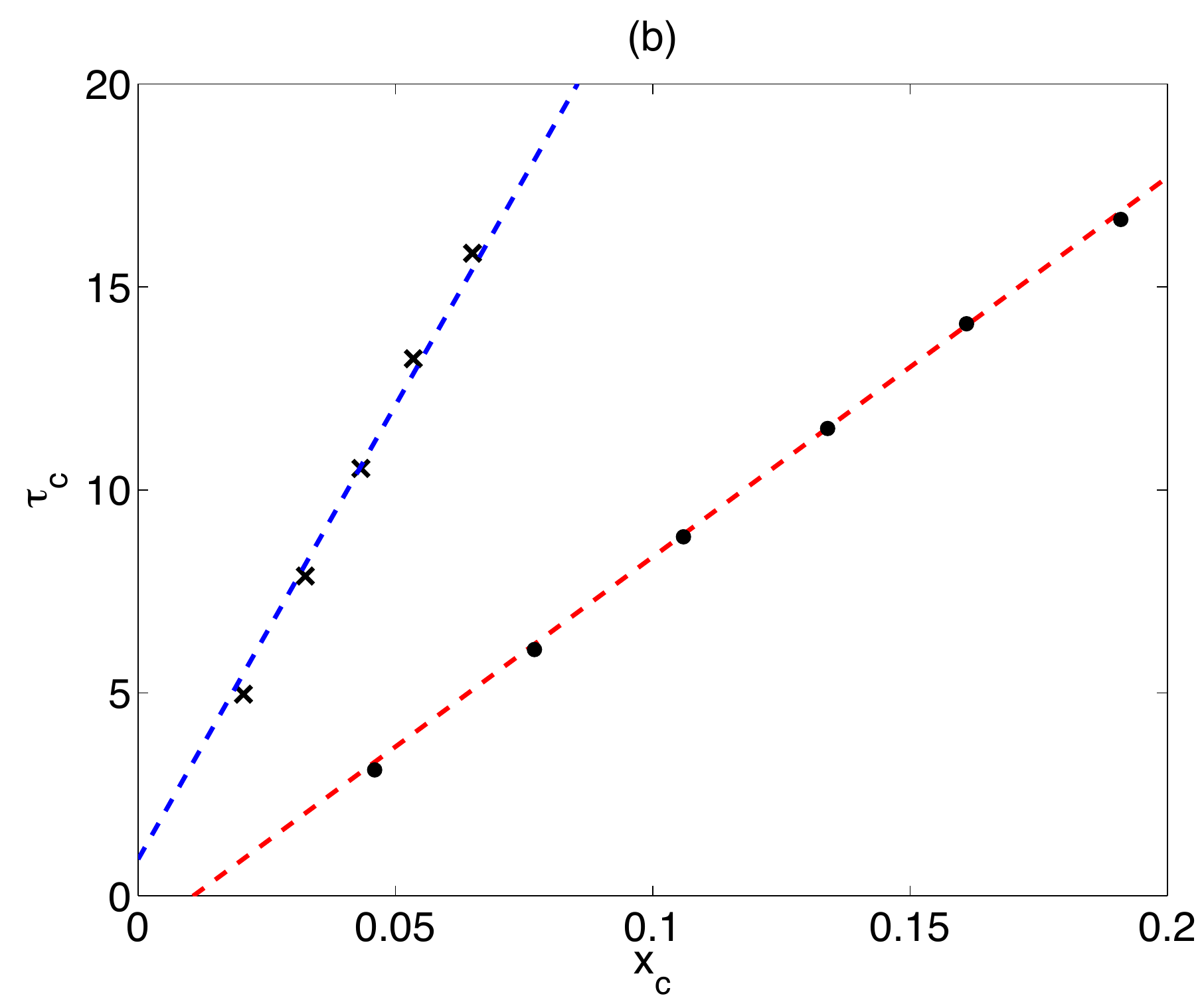}
\caption{
\label{fig_tc_chi}
Crossover time for transition to front acceleration in the fractional Fisher-Kolmogorov model in the presence of diffusion according to Eq.~(\ref{fk_frac_diff}) with  $\alpha=1.5$,  $\chi_d=5 \times 10^{-6}$ and $\gamma=1$, with initial condition in Eq.~(\ref{ic}).
The dots (crosses) denote the numerical values with initial condition in Eq.~(\ref{ic}) with $\nu=100$ ($\nu=1000$). Panel (a) shows $\tau_c$ as function of the fractional diffusivity $\chi_f$, and the dashed lines the logarithmic fit in Eq.~(\ref{cross_chi}). Panel (b) shows  $\tau_c$ as function of $x_c$, with the blue (red) dashed line corresponding to the linear fit, $x_c=v \tau_c$,  with $v=c_m$ ($v=c$) in Eq.~(\ref{c_min}) (Eq.~(\ref{c_diff})). }
%\vspace{10 cm}
\end{figure}

To quantify these ideas, we define the crossover spatio-temporal scale $(x_c,\tau_c)$ as the 
intersection of the Lagrangian diffusive front path in Eq.~(\ref{xlag_diff}),
with the Lagrangian fractional diffusion front path in Eq.~(\ref{lagx}). 
These intersection points are shown in the contour plots in Fig.~\ref{fig_fk_frac_diff}. 
Figure~\ref{fig_tc_chi}-(a)
shows $\tau_c$ as function of $\chi_f$ in semi-logarithmic scale for
$\alpha=1.5$,  $\chi_d=5 \times 10^{-6}$, and $\gamma=1$.  
The dots (crosses) correspond to a wide (narrow) front initial condition of the form in Eq.~(\ref{ic}) with 
$\nu=100$ ($\nu=1000$). 
It is observed that in both cases the crossover time can be fitted by a straight line which in the semi-logarithmic scale implies a scaling of the form 
\bq
\label{cross_chi}
\tau_{c} \sim \log \left( \chi_f^{-1} \right)\, ,
%t_{c} \sim \log \left( 1/ \chi_f \right)  \, ,
\eq
for 
$\alpha$,  $\chi_d$, and $\gamma$ fixed. 
Figure~\ref{fig_tc_chi}-(b)
shows $x_c$ as function of  $\chi_f$. As before, the results for $\nu=100$ and $\nu=1000$ can be fitted with straight lines indicating a dependence of the form 
\bq
\label{cross_x_chi}
x_c \sim  v \log \left( \chi_f^{-1} \right)\, .
\eq
However in this case the proportionality constant is different. For 
$\nu=100$ (dots), $v$ is the diffusive speed of wide fronts in Eq.~(\ref{c_diff}), and for 
$\nu=1000$ (crosses), $v$ is the diffusive minimum speed of narrow fronts in Eq.~(\ref{c_min}).

Using the scaling relation
\bq
_{-\infty} D_x^\alpha f \left(\zeta z\right) = \zeta^\alpha\,  _{-\infty} D_{\zeta x}^\alpha f \left(\zeta z\right) \, ,
\eq
where $\zeta$ is a constant, Eq.~(\ref{fk_frac_diff}) can be written in dimensionless form as
\bq
\label{dim_fk_frac_diff}
\partial_\tau \phi =\partial_z^2 \phi + \kappa \,  _{-\infty}D_x^\alpha \phi +  \phi \left ( 1 - 
\phi \right) \, ,
\eq
where
\bq
\tau= \gamma t \, , \qquad 
z=\sqrt{\frac{\gamma}{\chi_d}}\, x \, , \qquad
\kappa= \frac{\chi_f}{\gamma} \left(\frac{\gamma}{\chi_d}\right)^{\alpha/2} \, . 
\eq
In terms of dimensionless variables, the scaling in Eq.~(\ref{cross_chi}) implies 
$\tau \sim \log \kappa^{-1}$, that is
\bq
\tau_c \sim \frac{1}{\gamma} \log \left[ 
\frac{\gamma}{\chi_f}
\left(
\frac{\chi_d}{\gamma}
\right )^{\alpha/2}
\right]
\eq

\begin{figure}
\includegraphics[scale=0.4]{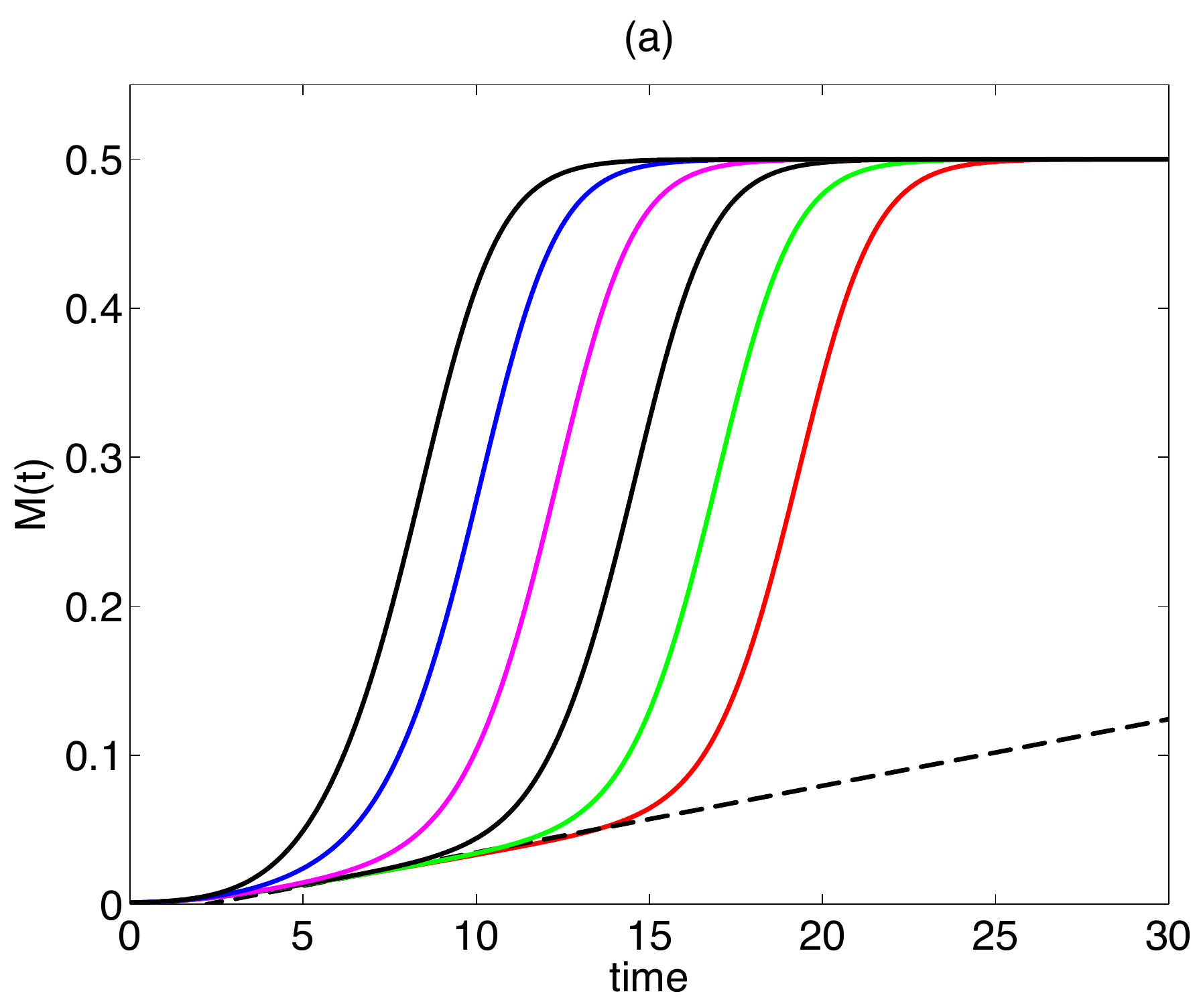}
\includegraphics[scale=0.4]{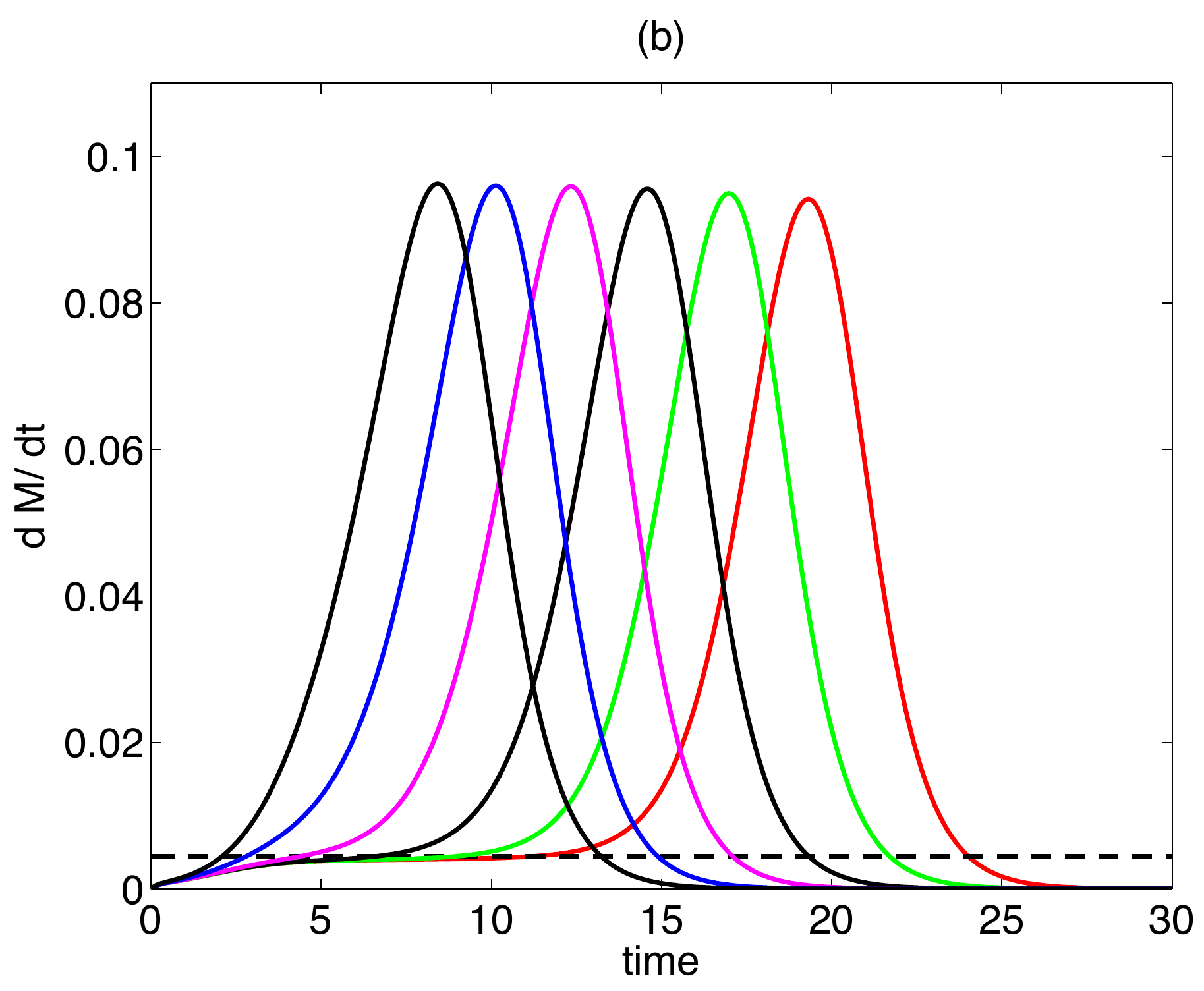}
\caption{
\label{fig_mass}
Time evolution of total concentration, $M(t;x_0)$ in Eq.~(\ref{mass}), and
rate of change of total concentration, $dM/dt$, for $x_0=0.5$ in  the fractional Fisher-Kolmogorov model in the presence of diffusion
according to Eq.~(\ref{fk_frac_diff}) with  $\alpha=1.5$,  $\chi_d=5 \times 10^{-6}$ and $\gamma=1$.
The different curves from left to right correspond to 
$\chi_f=5 \times 10^{-4},\, 5 \times 10^{-5},\, 5 \times 10^{-6},\, 5 \times 10^{-7},\, 5 \times 10^{-8}$,
and  $5 \times 10^{-9}$. The black dashed lines (a) and (b) 
show $M=c_m t$, and $dM/dt=c_m$ respectively with $c_m$ in Eq.~(\ref{c_diff}).
}
%\vspace{10 cm}
\end{figure}

A complementary way to quantify the dependence of the front dynamics on the ratio $\chi_f/ \chi_d$ is by considering the time evolution of the cumulative concentration, 
\bq
\label{mass}
M(t; x_0)=\int_0^{x_0}\phi(x,t)\, d x \, .
\eq
Figure~\ref{fig_mass} shows $M(t; x_0)$ and $dM(t; x_0)/dt$ 
for the narrow ($\nu=1000$) front case in Fig.~\ref{fig_tc_chi}.  
When only diffusive transport is present, i.e. when $\chi_f=0$, narrow fronts propagate at the constant minimum speed $c_m$ in Eq.~(\ref{c_min}), and thus $M(t; x_0) \sim c_m t$.  
Figure.~\ref{fig_mass} shows that this scaling describes the transient phase of the evolution of  
$M(t; x_0)$. Consistent with the previous results,  the time span of this transient phase increases with decreasing $\chi_f/ \chi_d$. Figure~\ref{fig_mass}-(b) shows $dM(t; x_0)/dt$, and it is observed that 
as  $\chi_f/ \chi_d$ decreases this function reaches its maximum at longer times, which is another manifestation of the delay of front acceleration induced by diffusive transport mentioned before.

%%%%%%%%%%%%%%%%%%%%%%%%%%%%%%%%%%%%%%%%%%%%
% FIGURE
%%%%%%%%%%%%%%%%%%%%%%%%%%%%%%%%%%%%%%%%%%%%
%%%%%%%%%%%%%%%%%%%%%%%%%%%%%%%%%%%%%%%%%%%%

%%%%%%%%%%%%%%%%%%%%%%%%%%%%%%%%%%%%%%%%%%%%
% FIGURE
%%%%%%%%%%%%%%%%%%%%%%%%%%%%%%%%%%%%%%%%%%%%
%%%%%%%%%%%%%%%%%%%%%%%%%%%%%%%%%%%%%%%%%%%%

%%%%%%%%%%%%%%%%%%%%%%%%%%%%%%%%%%%%%%%%%%%%
% FIGURE
%%%%%%%%%%%%%%%%%%%%%%%%%%%%%%%%%%%%%%%%%%%%
%%%%%%%%%%%%%%%%%%%%%%%%%%%%%%%%%%%%%%%%%%%%

%%%%%%%%%%%%%%%%%%%%%%%%%%%%%%%%
\section{Summary and conclusions}
%%%%%%%%%%%%%%%%%%%%%%%%%%%%%%%%

We have presented a study of the role of anomalous transport in front propagation in reaction-diffusion systems. Going beyond standard diffusion, we considered three models of anomalous transport: a fractional diffusion model, a tempered fractional diffusion model, and a combined standard and fractional diffusion model. The fractional diffusion model was based on the use of fractional derivatives which are non-local operators with algebraic decaying kernels. From the statistical mechanics perspective, fractional diffusion arises from the fluid limit of  continuous time random walk models with L\'evy flights. 
The characteristic signature in this case  is the appearance of algebraic decaying tails in the propagator (Green's function) which correspond to  $\alpha$-stable L\'evy distributions.  The main motivation behind the second anomalous transport model based on tempered fractional diffusion, is  the fact that L\'evy flights, i.e. random displacements with infinite second moments are a mathematical idealization. Systems of practical interest typically exhibit a cut-off scale beyond which L\'evy flight cannot extend. In this case, the second moment exhibits a super-diffusive scaling in an intermediate asymptotic regime, followed by a diffusive scaling in the asymptotic limit.
In this case the 
propagator which correspond to tempered L\'evy distribution exhibits algebraic decay in an intermediate range but the asymptotic decay of the tail is exponential. The study of the third transport model based on a combination of fractional and regular diffusion is also motivated by practical considerations. In particular, it is natural to expect that in addition to fractional diffusion some irreducible remnant of regular diffusion to be present on transport. The Green's function for the combined operator exhibits a Gaussian-type dependence near the origin followed by an $\alpha$-stable L\'evy-type dependence at large distances. 

For the reaction kinetics we consider the Fisher-Kolmogorov nonlinearity that has one stable and one unstable steady state. In the case of normal transport, the interaction  of the reaction kinetics with diffusion leads to the propagation of fronts in which the stable phase invades the unstable phase.
The two key aspect of the dynamics in this  case is the constant speed of the propagation and the exponential decay of the front's tail.  Our main goal was to present a numerical study of how this standard well-understood front dynamics is modified in the presence of anomalous transport. The numerical method was based on an operator splitting algorithm with an explicit Euler step for the time advance of the reaction kinetics and a Crank-Nicholson semi-implicit time step for the transport operator. In the case of regular diffusion, the transport operator was discretize using a centered finite-difference method. In the cases of fractional  diffusion and fractional tempered diffusion we use a flux-conserving 
scheme with an upwind Grunwald-Letnikov finite-difference discretization of the regularized fractional derivative. 

The main differences found in the transport scenarios considered, concern the propagation speed of the fronts, and the 
spatial decay of the tails of the fronts. As it is well-known, in the case of regular diffusion, fronts propagate at a constant speed and the tails decay exponentially fast. 
However, in the presence of fractional diffusion of order $\alpha$, fronts exhibit an exponential Lagrangian acceleration, $a_L(t) \sim e^{\gamma t/\alpha}$, and develop algebraic decaying tails, $\phi \sim 1/x^{\alpha}$.  In the case of tempered fractional diffusion, this phenomenology prevails in the intermediate asymptotic regime
 $\left(\chi_\tau t \right)^{1/\alpha} \ll x \ll 1/\lambda$, where $1/\lambda$ is the scale of the exponential tempering. Outside this regime, i.e. for $x > 1/\lambda$, the tail of the front exhibits the tempted decay 
$\phi \sim e^{-\lambda x}/x^{\alpha+1}$, the acceleration is transient and the Lagrangian front velocity approach the terminal speed $v_L \rightarrow v_*= \left(\gamma-\lambda^\alpha \chi\right)/ \lambda$. When fractional and regular diffusion are both present, the fronts also accelerate. However, the onset of the acceleration is retarded by the regular diffusion. 
In particular, the numerical results show that the crossover time time, $t_c$, to transition to the accelerated fractional regime exhibits a logarithmic scaling of the form $t_c \sim \log \left(\chi_d/\chi_f\right)$. Regular diffusion also modifies the algebraic spatial decay of the tails observed in the presence of pure  fractional diffusion. 
%There is also a cross-over in the transition from exponential decay to algebraic decay in the spatial decay of the tails of the fronts.  
In particular, the crossover scale, $x_c$, to transition to the algebraic decay of the front's tail is given by $x_c= v t_c$ where $v$ is the diffusive front speed. The diffusive delay was also quantified using  the  cumulative concentration function, $M$, and its rate of change, $dM/dt$. It was observed that as  $\chi_f/ \chi_d$ decreases, $dM/dt$, reaches its maximum at longer times. 

 % %%%%%%%%%%%%%%%%%%%%%%%%%
 \section{Acknowledgments}
 % %%%%%%%%%%%%%%%%%%%%%%%%%
This work was sponsored by the Office of Fusion Energy Sciences 
of the US Department of 
Energy at Oak Ridge National Laboratory, managed by UT-Battelle, LLC, 
for the U.S.Department of Energy under contract DE-AC05-00OR22725.

%%%%%%%%%%%%
 % %% Bibliography
 % %

\end{document}